\documentclass[preprint2]{aastex}
\usepackage{amsmath}

\shorttitle{Variability of H$\beta$ line profiles as an indicator of orbiting bright spots}
\shortauthors{Jovanovi\'{c} et al.}

\begin{document}

\title{Variability of the H$\beta$ line profiles as an indicator of orbiting bright spots
in accretion disks of quasars: a case study of 3C 390.3}

\author{P. Jovanovi\'{c}\altaffilmark{1,2}, L. \v{C}.
Popovi\'{c}\altaffilmark{1,2}, M. Stalevski\altaffilmark{1,2,3}}
\affil{\altaffilmark{1}Astronomical Observatory, Volgina 7, 11060 Belgrade,
Serbia}
\affil{\altaffilmark{2}Isaac Newton Institute of Chile, Yugoslavia Branch}
\affil{\altaffilmark{3}Sterrenkundig Observatorium, Universiteit Gent,
Krijgslaan 281-S9, Gent,
9000, Belgium}
\email{pjovanovic@aob.rs, lpopovic@aob.rs, mstalevski@aob.rs}
\and
\author{A. I. Shapovalova}
\affil{Special Astrophysical Observatory of the Russian AS, Nizhnij
Arkhyz, Karachaevo-Cherkesia 369167, Russia} \email{ashap@sao.ru}

\begin{abstract}

Here we show that in the case when double peaked emission lines
originate from outer parts of accretion disk, their variability
could be caused by perturbations in the disk emissivity.  In order
to test this hypothesis, we introduced a model of disk perturbing
region in the form of a single bright spot (or flare) by a
modification of the power law disk emissivity in appropriate way.
The disk emission was then analyzed using numerical simulations
based on ray-tracing method in Kerr metric and the corresponding
simulated line profiles were obtained. We applied this model to the
observed H$\beta$ line profiles of 3C 390.3 (observed in the period
1995-1999), and estimated the parameters of both, accretion disk and
perturbing region. Our results show that two large amplitude
outbursts of the H$\beta$ line observed in 3C 390.3 could be
explained by successive occurrences of two bright spots on
approaching side of the disk. These bright spots are either
moving, originating in the inner regions of the disk and spiralling
outwards by crossing small distances during the period of several
years, or stationary. In both cases, their widths increase with
time, indicating that they most likely decay.

\end{abstract}

\keywords{galaxies: active --- quasars: individual (3C 390.3) ---
quasars: emission lines --- line: profiles}

\section{Introduction}

The huge amount of Active Galactic Nucleus (AGN) energy is released through
accretion onto super-massive black hole (BH), supposed to exist in the center of
AGN. The emission of the accretion disk is not only in the continuum, but
also in the emission lines (e.g. in Fe K$\alpha$ line) and in low
ionization lines, as e.g. in broad Balmer emission lines which are seen as double
peaked (DP). DP Balmer lines  are found in 20\% of radio loud AGN  at $z <
0.4$ \citep{eh94,eh03} and 4\% of the Sloan digital Sky Survey (SDSS) quasars
at $z < 0.33$ \citep{st03}.

Broad, double-peaked emission lines of AGN
provide dynamical evidence for presence of an accretion disk feeding
a supermassive black hole in the center of AGN. But in some cases,
the variability of these lines shows certain irregularities which
could not be explained just by standard model of accretion disk.

The DP line profiles are often used to extract the disk parameters
\citep[see e.g.][]{ch89,ch90,eh94,eh03,st08,er09}.
In a series of papers Dumont \& Collin-Souffrin \citep[see][and references
therein]{cs87,csd90,dcs90a,dcs90b,dcs90c}
investigated the radial structure and emission of the outer regions of the optically thin accretion disks in AGN
and calculated detailed grid of photoionisation models in order to predict the relative strengths of low-ionization
lines emitted from the disk. They obtained integrated line intensities and line profiles emitted at each radius of
the disk, for its different physical parameters. They also studied the influence of the external illumination
on the structure of the disk, considering the point source model, where a compact source of non-thermal radiation
located at a given height illuminates the disk and the diffusion model, where the radiation of a central source is
scattered back towards the disk by a hot diffusing medium. One of the first methods for calculating the profiles of
optical emission lines from a relativistic accretion disk was proposed by \citet{ch89}. The limitation to this method
is that the accretion disk structure required to
explain the variability of the line profiles cannot be axi-symmetric, i.e.
very often the red peak is higher than blue one and that cannot be explained
by this model. It is  not possible in a circular disk, in which the blue
peak is always Doppler boosted to be stronger than the red peak.
Therefore, \citet{er95} adapted the circular accretion disk
model to elliptical disks in order to fit the profiles of double-peaked
emitters with a red peak stronger than the blue one. This model
introduced eccentricity and phase angle parameters to the circular model
described above, and  the pericenter distance of the
elliptical orbits \citep[see][]{er95}.

Spectroscopic monitoring of double-peaked emitters \citep[see
e.g.][]{sh01,ge07,sh09}
has revealed that a ubiquitous property of the double-peaked broad emission
lines is variability of their profile shapes on the timescales of months to years. DP
line profiles are observed to vary on timescales of months to years, i.e. on
timescales of the order of the dynamical time or longer
\citep[e.g.][]{vz91,zh91,ma93,ro98,se00,sh01,sb03,ge07}.

This slow, systematic variability of the line profile is on the timescale
of dynamical changes in an accretion disk, and has been shown to be
unrelated to the shorter timescale variability  seen in the
overall flux in the line, due to reverberation of the variable ionizing
continuum. Patterns in the variability of the broad Balmer lines are
often a gradual change and reversal of the relative strengths of the
blue-shifted and red-shifted peaks \citep[see e.g.][]{ne97}.

Periodic variability of the red and blue peak strengths has also been
attributed to a precessing elliptical disk, a precessing single-armed
spiral (as e.g. 3C 332, 3C 390.3: \citet{gi99}; NGC 1097: \citet{sb03}), and a
precessing warp in the disk. For instance, \citet{wu08} computed
the profiles of Balmer emission lines from a relativistic, warped accretion disk
in order to explore the certain asymmetries in the double-peaked
emission line profiles which cannot be explained by a circular Keplerian disk.
Elliptical disks and spiral waves have been
detected in cataclysmic variables \citep{shh97,bc00}, and
a radiation induced warp has been detected in the large-scale disk of the AGN
NGC 4258 \citep{mbp96}.

Spiral waves are a physically desirable model since they can be produced
by instability in the vicinity of a black hole. They can play an important
role in accretion disks, because they provide a mechanism for transporting
angular momentum outward in the disk, allowing the gas to flow inwards, towards
the central black hole. Long-term profile variability is thus a useful tool for
extracting information about the structure and dynamics of the accretion disk
most likely producing the double-peaked emission lines.

In this paper, we
present an investigation of the disk line variations due to instability in accretion
disk. First we developed a model, assuming that instability in the accretion
disk affects disk emissivity. This model and some simulations of expected line
profile variability are presented in \S 2. In \S 3 we compare the model with
observations taken from long-term monitoring of 3C 390.3 \citep{sh01} in order to
obtain parameters of perturbations. In \S 4 we discuss our results in the light
of possible physical mechanisms which could cause such perturbations, and
finally, in \S 5 we outline our conclusions.

\section{The model of perturbation in the accretion disk}

Here we introduce the model and some approximations used in the simulations of
accretion disk perturbation.

\subsection{Long term variation of DP line profiles: some assumptions and
problems }

As we mentioned above, the DP line profile variability does not
appear to correlate with changes in the line and/or continuum flux,
and consequently one can assume that the changes in line profile are
likely caused by changes in the accretion disk structure. There are
several examples of the long-term variability (on timescales
of several years) of the DP line profile of some objects which has
been successfully modeled by the precession of a non-axisymmetric
accretion disk, such as an elliptical disk or a disk with a spiral
arm \citep[][and references therein]{ge07,sb03,sh01,gi99}. These
models, however, fail to explain the long-term variability
of some objects and the short-term variability (on timescales from
several months to a year) of all objects \citep{le05,le10}. For
instance, \citet{le10} found that the two simple models, an
elliptical accretion disk and a circular disk with a spiral arm, are
unable to reproduce all aspects of the observed variability,
although both account for some of the observed behaviors. Therefore,
these authors suggest that many of the observed variability patterns
could be reproduced assuming a disk with one or more fragmented
spiral arms.

Other attempts to explain the DP line profile variability through
perturbations of the disk structure introduced bright spots over an
axisymmetric accretion disk. As an example  \citet{ne97}
successfully modeled the variation of the H$\alpha$ peak intensity ratio of Arp 102B
with a single spot rotating within the disk, but \citet{ge07} were
not able to apply the same model to the same object at a different time period.

In the case of Fe K$\alpha$ variability \citet{tu06} used the spot model to
explain the variability of the iron line profile of Mrk 766 in the X-ray band.
Also, \citet{do08} studied variations of the iron line due to an orbiting spot
which arise by reflection on the surface of an accretion disk, following its
illumination by an X-ray flare in form of an off-axis point-like source just
above the accretion disk. Besides the spots in accretion disk, the Fe K$\alpha$
line of some AGN could be also significantly affected by highly-ionized fast
accretion disk outflows. For instance, \citet{sm10} found that the major
features in the observed 2 -- 10 keV spectrum of the bright quasar PG1211+143
can be well reproduced by their Monte Carlo radiative transfer simulations
which include a variety of disk wind (outflow) models.

To explain the short-timescale variability of the DP line profiles
\citet{fe08} constructed stochastically perturbed accretion disk
models and calculated H$\alpha$ line profile series as the bright spots
rotate, shear and decay. They  ruled out spot production by star/disk
collisions and favor a scenario where the radius of marginal self-gravity
is within the line emitting region, creating a sharp increase in the
radial spot distribution in the outer parts.

\subsection{The model of bright spot--like perturbing region}

We model the emission from accretion disk using numerical simulations based on
ray-tracing method in Kerr metric \citep[see e.g.][and references therein]{pj09a}.
Although this method was developed for studying the X-ray radiation which originates from the
inner parts of the disk close to the central black hole \citep[see e.g.][]{pj08a},
it can be also successfully applied
for modelling the UV/optical emission which originates from the outer regions of the
disk\footnote{As shown in \citet{pj08a}, the effects of strong gravitational field and
angular momentum of rotating black hole are significant only in the innermost parts of
accretion disk, in vicinity of the central supermassive black hole, up to the several dozens
of gravitational radii. In the outer parts of the disk, such as those where H$\beta$ line originates,
these effects are negligible and Kerr metric with zero angular momentum, i.e. Schwarzschild metric,
is a very good approximation.}.

Surface emissivity of the disk is usually assumed to vary with
radius as a power law \citep[e.g.][]{lcp03}: $\varepsilon (r) =
\varepsilon _0  \cdot r^q,$ where $\varepsilon _0$ is an emissivity
constant and $q$ -- emissivity index. Total observed flux is then
given by:
\footnotesize
\begin{equation}
F_{obs} (E_{obs} ) = \int\limits_{image} {\varepsilon (r) \cdot g^4 e^{ - \dfrac{1}{2}\left(\dfrac{E_{obs}  -
gE_0}{\sigma} \right)^2}} d\Xi ,
\label{eq1}
\end{equation}
\normalsize
where $g$ is the energy shift due to the relativistic effects: $g =
\dfrac{{\nu _{obs} }}{{\nu _{em} }}$, $E_0$ is the rest energy of
the line, $\sigma$ is the local turbulent broadening and $d\Xi$ is the solid angle subtended by the disk in the
observer's sky.

\begin{figure}
\centering
\includegraphics[width=0.4\textwidth]{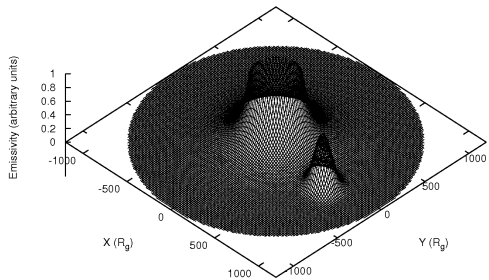}
\caption{A 3D plot of modified disk emissivity given by Eq. (\ref{eq2}) for
$200\ \mathrm{R_g}\le r(x,y) \le 1200\ \mathrm{R_g}$, $q=-2.5$
and for the following parameters of perturbing region: $\varepsilon_p=5$,
$x_p=700\ \mathrm{R_g}$, $y_p=-150\ \mathrm{R_g}$ and
$w_x = w_y = 100\ \mathrm{R_g}$.}
\label{fig1}
\end{figure}

In this paper we adopt the following modification of the power-law
disk emissivity, in order to introduce a bright spot--like perturbing
region in the disk \citep{pj08b,pj09a,pj09b,ms08}:
\footnotesize
\begin{equation}
\begin{array}{lll}
 \varepsilon_1 (x,y) &= \varepsilon (r(x,y))& \\ 
&\cdot \left( {1 + \varepsilon _p  
\cdot e^{ - \left( {\left( {\dfrac{{x -x_p}}{{w_x }}} \right)^2  + \left( {\dfrac{{y - y_p }}{{w_y }}} \right)^2 }
\right)} } \right),&
\end{array}
\label{eq2}
\end{equation}
\normalsize
where $\varepsilon_1 (x,y)$ is the modified disk emissivity at the
given position $(x,y)$ expressed in gravitational radii $R_g$,
$\varepsilon (r(x,y))$ is the ordinary power-law disk emissivity at
the same position, $\varepsilon_p$ is emissivity of the perturbing
region (i.e. amplitude of the bright spot), $(x_p,y_p)$ is the
position of perturbing region with respect to the disk center (in
$R_g$) and $(w_x, w_y)$ are its widths (also in $R_g$). A 3D plot of
above expression for modified emissivity law is given in Fig.
\ref{fig1}.
This simple model is suitable for our purpose because it
allows us to change amplitude, width and location of bright spot in
respect to the disk center. In that way we are able to simulate
displacement of bright spot along the disk, its widening and
amplitude decrease with time (decay). Moreover, the above bright spot model
can be successfully applied for studying the variations of accretion disk emission
in different spectral bands, from X-rays to optical band \citep[see e.g.][]{pj08b,pj09b,ms08}.

\section{Results: Model vs. observations}

\subsection{Perturbation in the accretion disk: modeled profiles}

\begin{figure*}
\centering
\includegraphics[width=0.49\textwidth]{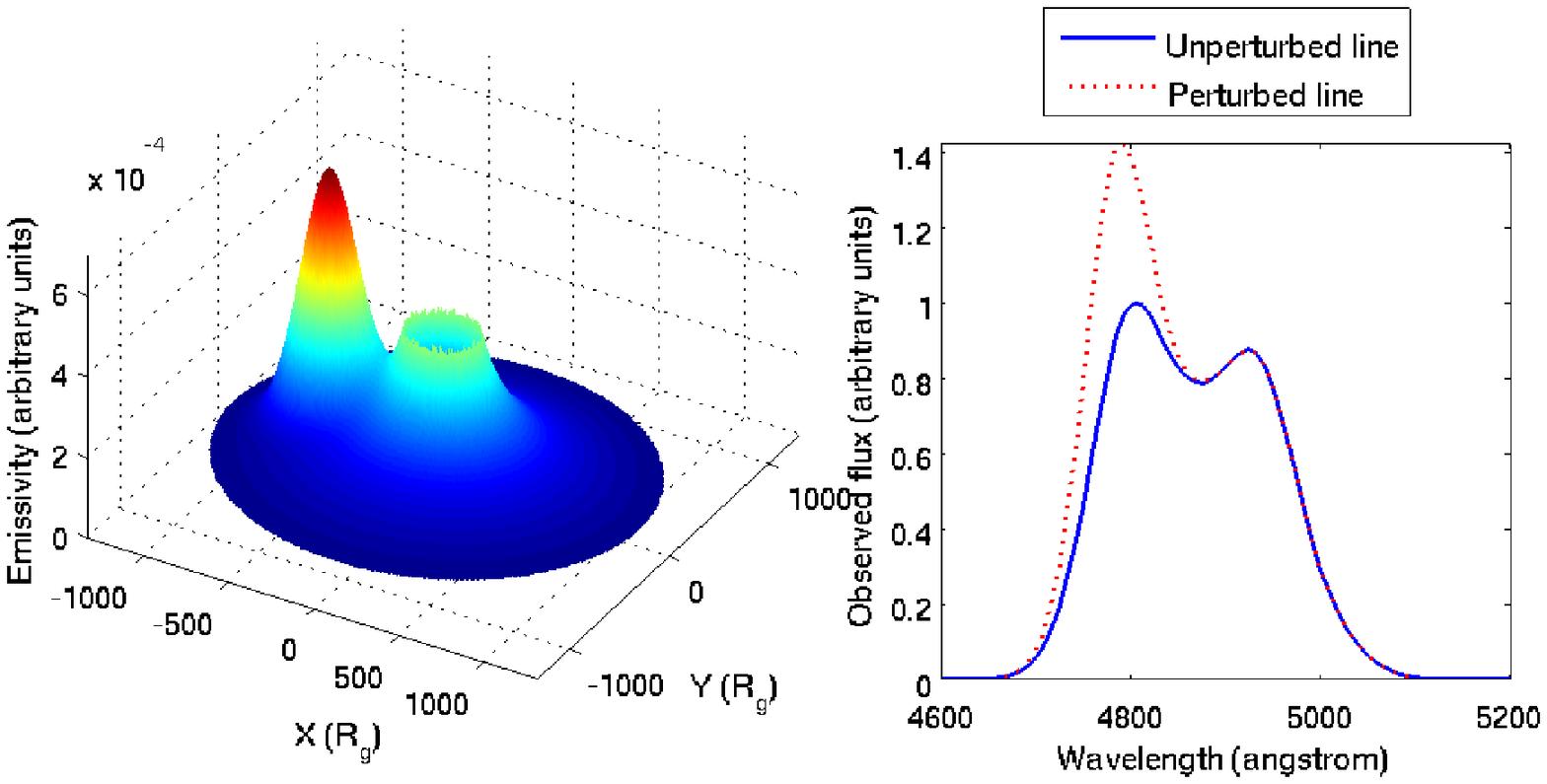}
\includegraphics[width=0.49\textwidth]{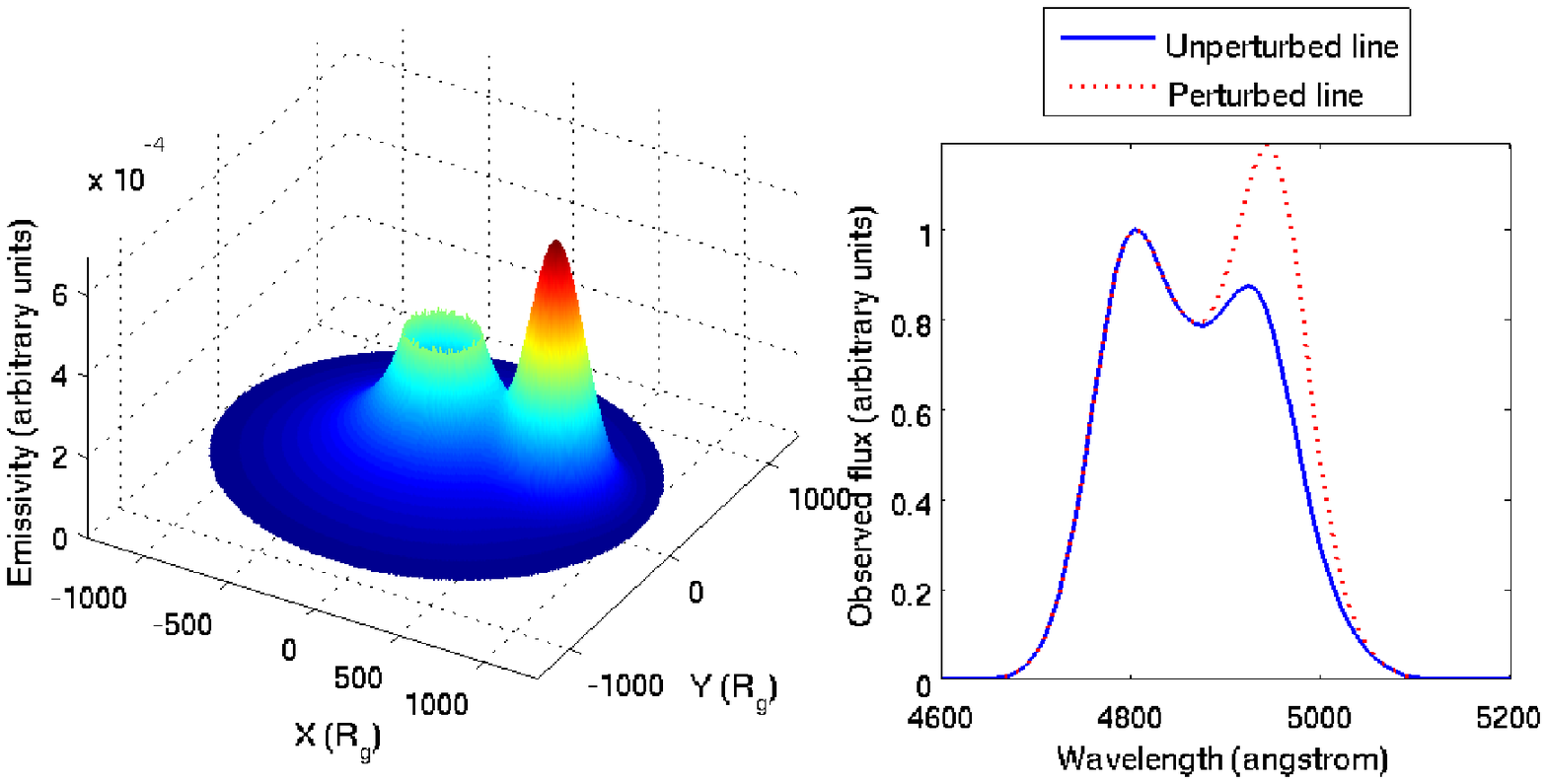} \\
\includegraphics[width=0.49\textwidth]{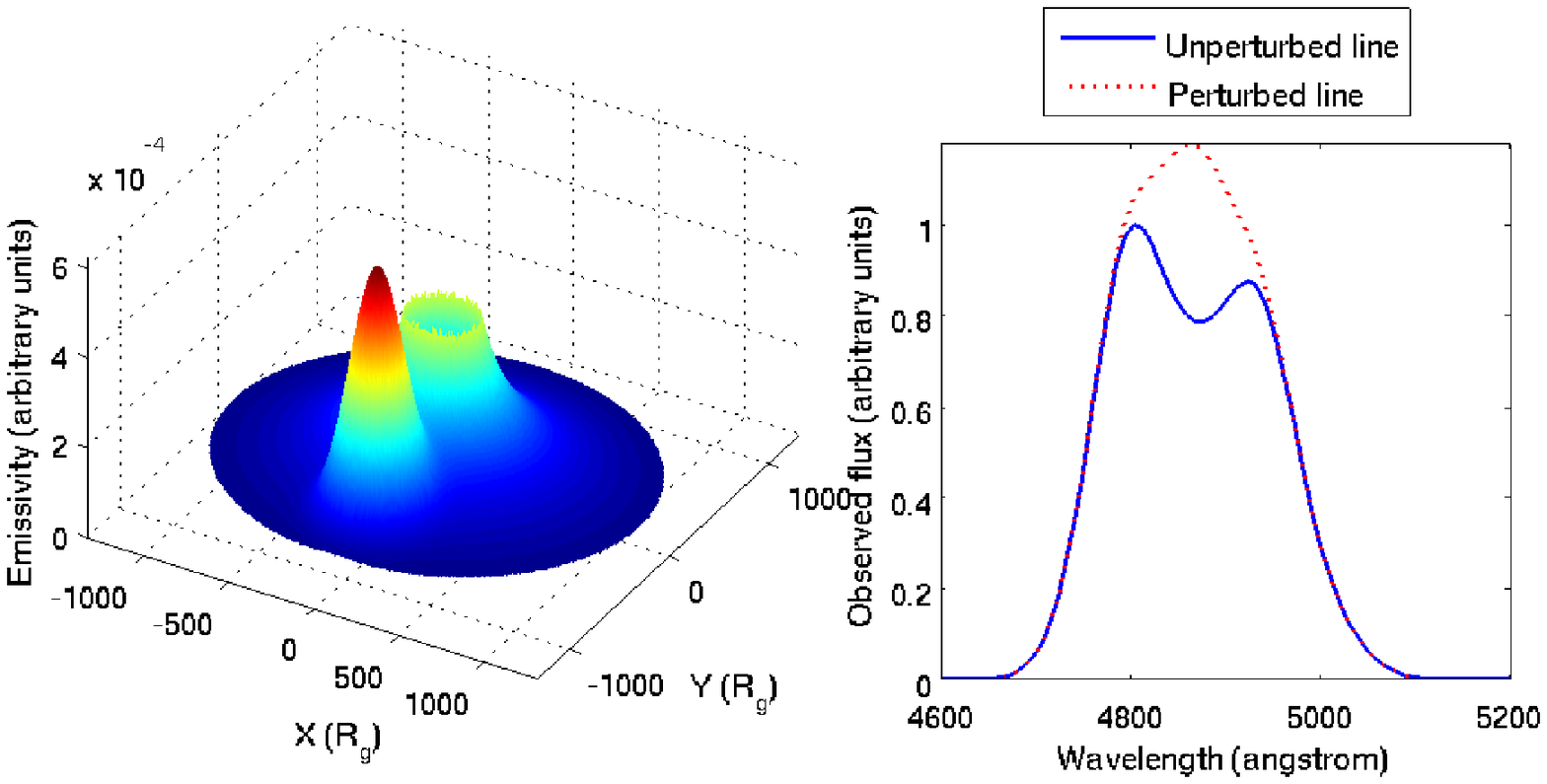}
\includegraphics[width=0.49\textwidth]{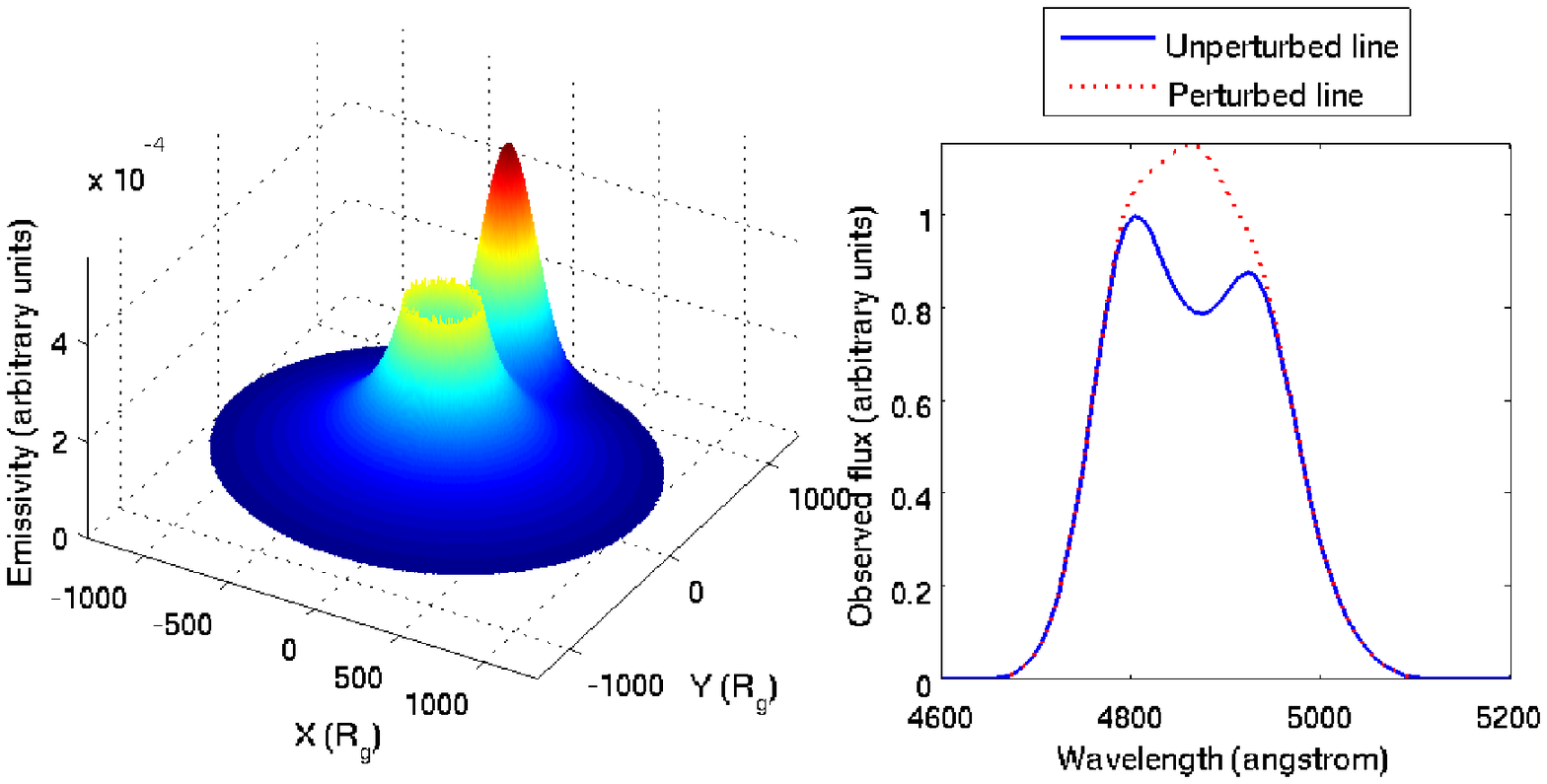}
\caption{Shapes of perturbed emissivity of an accretion
disk in Kerr metric and the corresponding perturbed (dashed line) and
unperturbed (solid line) H$\beta$ line profiles for the following parameters of
perturbing region:
$\varepsilon_p=5$, $w_x=w_y=200\ R_g$. The positions of perturbing region are:
$x_p=-700\ R_g$ and $y_p=0$ (top left), $x_p=700\ R_g$ and $y_p=0$ (top right),
$x_p=0$ and $y_p=-700\ R_g$ (bottom left) and $x_p=0$ and $y_p=700\ R_g$ (bottom
right).}
\label{fig2}
\end{figure*}

In order to test how this model of bright spot affects the H$\beta$
line profile, we performed several numerical simulations of
perturbed emission of an accretion disk in Kerr metric for
different positions of bright spot along $x$ and $y$-axes in both,
positive and negative directions. For these simulations we adopted the following
parameters of the disk: inclination $i=30^o$, inner and outer radii
$R_{in}=200$ and $R_{out}=1200\ R_g$, power law emissivity with index
$q=-1$, local turbulent broadening $\sigma=2000\ \mathrm{km\ s^{-1}}$ and
normalized
angular momentum of black hole $a=0.5$. The corresponding results are presented
in Fig. \ref{fig2}. As it can be seen from this figure, when the
bright spot moves along the positive direction of $x$-axis (receding
side of the disk) it affects only "red" wing of the line (Fig.
\ref{fig2}, top right), but when it moves along the negative direction of
$x$-axis (approaching side of the disk) it affects only "blue" wing
of the line (Fig. \ref{fig2}, top left). In both cases, the other wing and the
line core stay nearly constant, and therefore almost unaffected by
bright spot. The situation is quite opposite when the bright spot
moves in both directions along the $y$-axis, because then it affects
only the line core, while the both of its wings stay almost intact
(see the bottom panels of Fig. \ref{fig2}).

\begin{figure*}
\centering
\includegraphics[width=0.49\textwidth]{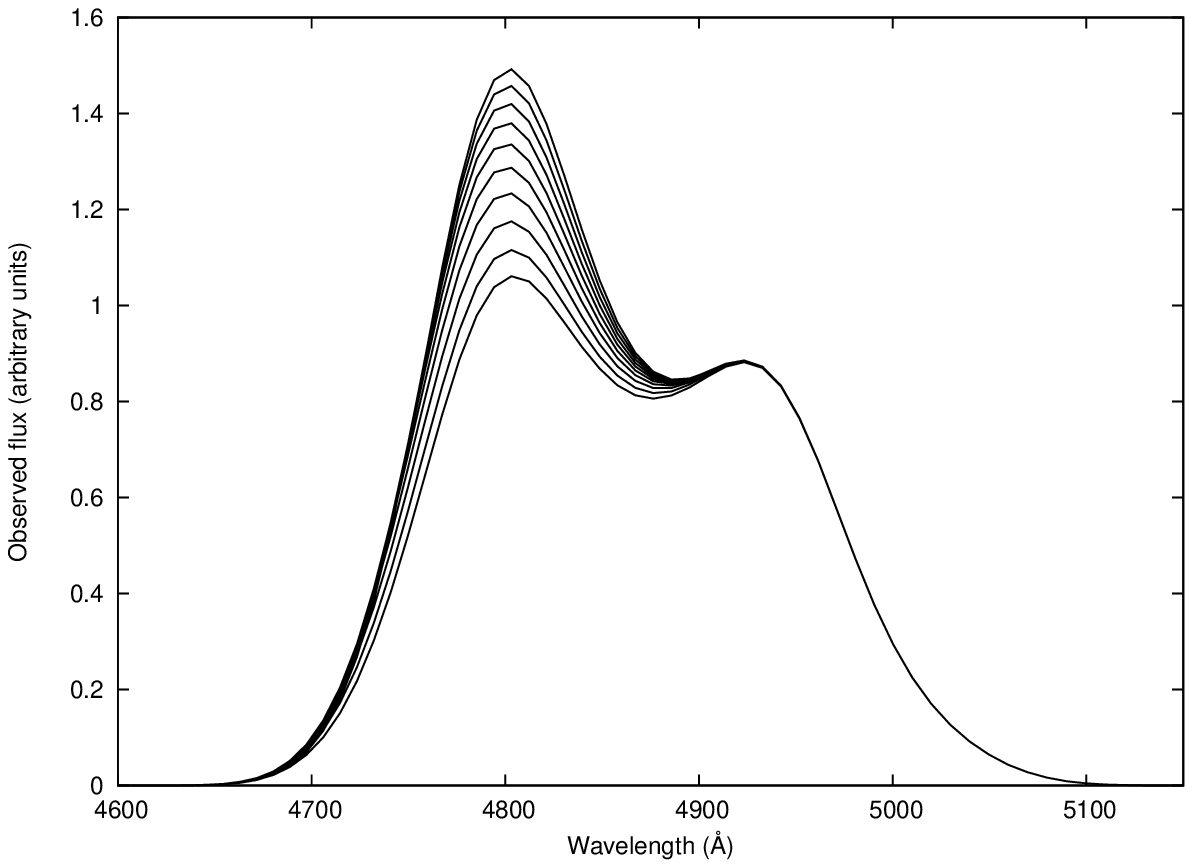}
\includegraphics[width=0.49\textwidth]{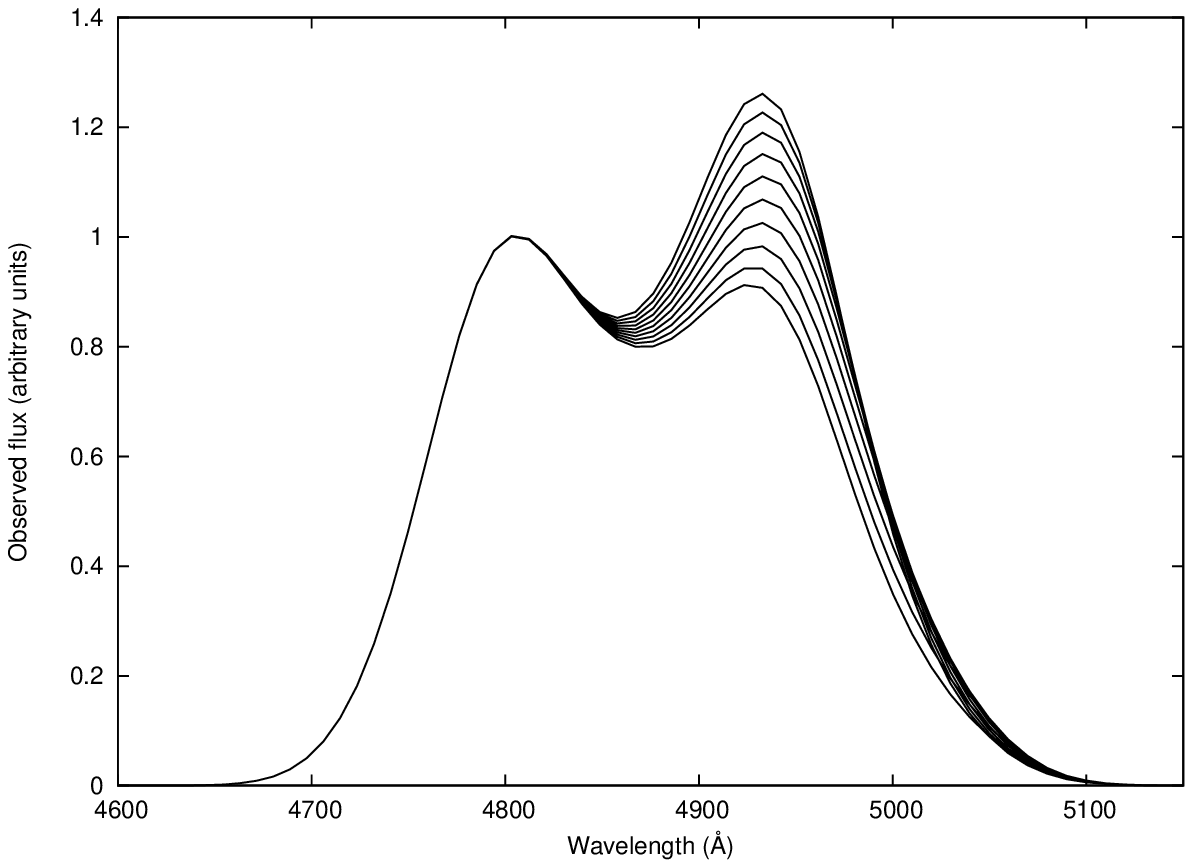}
\caption{Variations of the perturbed H$\beta$ line profile for different
positions of bright spot
along the $y=x$ direction. Disk parameters are the same as used in examples in
Fig. \ref{fig2}. Left panel corresponds to the positions of bright spot on
approaching side of the disk, while the right panel corresponds to the receding
side of the disk. In both cases, the positions of bright spot were varied from
the inner radius (bottom profiles) of the disk towards its outer radius
(top profiles).}
\label{fig3}
\end{figure*}

We also performed the corresponding simulations for different positions of a
bright spot which moves from the inner radius of the accretion disk towards its
outer parts along the $y=x$ direction, and found similar behavior of the
simulated line profiles (see Fig. \ref{fig3}). As one can see from Fig.
\ref{fig3}, for certain positions of perturbing region along $y=x$ direction we
obtained the line profile with almost symmetrical wings, while in other cases
either the "blue" peak is brighter than the "red" one, or the "red" peak is
stronger than the "blue" one.

The next step in our analysis was to use our numerical simulations for fitting the observed
spectra of 3C 390.3 in order to study the variability of its H$\beta$ spectral line
due to emissivity perturbations in its accretion disk.

\subsection{Observations of 3C 390.3}

To test the model,  we used 22 spectra of 3C 390.3 observed from
November 1995 until June 1999 \citep[see Fig. 8 in][]{sh01}.

Spectra of 3C 390.3  were taken with the 6 m and 1 m telescopes of the SAO RAS
(Russia, 1995--2001) and at INAOE's 2.1 m telescope of the ``Guillermo Haro
Observatory'' (GHO) at Cananea, Sonora, M\'exico (1998--1999) in monitoring
regime in 1995-1999. They were obtained with long slit spectrographs, equipped
with CCD detector arrays. The typical wavelength interval covered was that from
4000\,\AA\, to 7500\,\AA, the spectral resolution varied between 5 and 15 \AA\
and the S/N ratio was $>$ 50 in the continuum near H$\alpha$ and H$\beta$.
Spectrophotometric standard stars were observed every night.
The spectrophotometric data reduction was carried out either with software
developed at SAO RAS or with the IRAF package for the spectra obtained
in M\'exico. The image reduction process included bias, flat-field corrections,
cosmic ray removal, 2D wavelength linearization, sky spectrum subtraction,
addition of the spectra for every night, and  relative flux calibration
based on standard star observations. Spectra were scaled by the
[\ion{O}{3}] $\lambda\lambda$4959+5007 integrated line flux
under the assumption that the latter did not change during the time interval
covered by our observations (1995--2001). A value of 1.7$\times$10$^{-13}$ ergs
s$^{-1}$cm$^{-2}$
\citep{vz91} for the integrated [\ion{O}{3}] line
flux was adopted.
In order to calculate a normalization coefficient,
the continuum was determined in two 30\,\AA\, wide clean --line free-- windows
centered at 4800\,\AA\, and 5420\,\AA. After continuum subtraction,
blend separation of the H$\beta$ and [\ion{O}{3}] components was carried out
by means of a Gaussian fitting procedure, applied to the following:
H$\beta$ --- broad blue, broad red and central narrow;
[\ion{O}{3}] $\lambda\lambda$4959,5007 --- broad and narrow components.
The forbidden lines are represented by two Gaussian curves with an intensity
ratio I(5007)/I(4959)=2.96.

Comparisons between mean and rms spectra of the H$\alpha$ and H$\beta$ broad line profiles of 3C 390.3
during 1995-2007 (including the H$\beta$ spectra from this paper) are given in
Fig. 12 of \citet{sh10}. It can be seen that these profiles are similar, and
moreover, the corresponding H$\alpha$ profiles of 3C 390.3 from \citet{sh10} and \citet{ge07} are also similar.
These comparisons indicate that the [\ion{O}{3}] $\lambda\lambda$4959,5007 narrow lines subtraction
in H$\beta$ region was performed correctly.

The flux of H$\beta$ and the broad component profile was obtained from scaled
spectra after continuum subtraction and removal of the [\ion{O}{3}]
doublet and narrow H$\beta$ component.
Then the observed continuum fluxes  were corrected for the aperture
effects using scheme by \citet{pe95}.
The mean error (uncertainty) in our flux determinations
for the continuum and H$\beta$ flux is $<$3\%. More details can be found
in paper \citet{sh01}.

\subsection{Perturbation in the accretion disk of 3C 390.3}

Radial velocities of the blue and red peaks of the H$\beta$ and H$\alpha$ broad lines of 3C 390.3 vary
with time \citep[see e.g.][]{ga96,er97,sh10}. \citet{sh01}
obtained the H$\beta$ difference profiles by subtracting the average spectrum
corresponding to the minimum activity state (September 9, 1997) from the individual spectra (see their Fig. 11).
These authors found (see their Table 7) that the radial velocity of the blue peak
increased from -3200 km s$^{-1}$ in 1995-1996 to -5200 km s$^{-1}$ in 1999. At the same time the radial velocity
of the red peak increased from +4900 km s$^{-1}$ in 1995-1996 to +7000 km s$^{-1}$ in 1999.
Here we analyze the possibility that the velocities, corresponding to the peak shifts in the H$\beta$ integral and
difference profiles, vary with time due to perturbations in disk emissivity.

In order to fit the spectral H$\beta$ line shapes of 3C 390.3, we
first estimated the disk parameters from several profiles. We found
following parameters  of the disk: inclination $i=20^o$, inner and
outer radii $R_{in}=100$ and $R_{out}=1300\ R_g$, broken power law
emissivity with index $q=-1$ for $R_{in}<r<R_{br}$ and $q=-3$ for
$R_{br}<r<R_{out}$, radius at which slope of emissivity changes
$R_{br}=500\ R_g$, emissivity of perturbing region $\varepsilon
_p=1$, local turbulent broadening $\sigma=2000\ \mathrm{km\ s^{-1}}$ and
normalized angular momentum of black hole $a=0.5$.
These values of parameters are in
accordance with the corresponding parameters for 3C 390.3 obtained by
\citet{fe08}, who found the following values: $i=27^o$, $R_{in}=450$,
$R_{out}=1400\ R_g$ and $\sigma=1300\ \mathrm{km\ s^{-1}}$.
In all our simulations we held emissivity of perturbing region fixed at $\varepsilon
_p=1$, but nevertheless, the brightness of perturbing region varies with time due to its changeable area
(i.e. its widths are taken as free parameters which vary with time), as well as due to power law emissivity
of the disk which decreases with radius.
We studied stationary and moving perturbing regions, and in the latter case their coordinates are also assumed as free
parameters.

The fitting of the observed H$\beta$ line shapes of 3C 390.3 is performed separately in the case of
moving and stationary perturbing regions in the following three steps: (i) the simulated H$\beta$ line
profiles are calculated for some initial set of values of free parameters; (ii) the corresponding
root mean square (RMS) of residuals between the observed and simulated line profiles is calculated;
(iii) the values of free parameters are varied and the procedure (i)-(iii) is repeated until the
RMS deviations (RMSD) between the observed and fitted H$\beta$ profiles become
as small as possible.

\begin{figure*}
\centering
\includegraphics[width=0.32\textwidth]{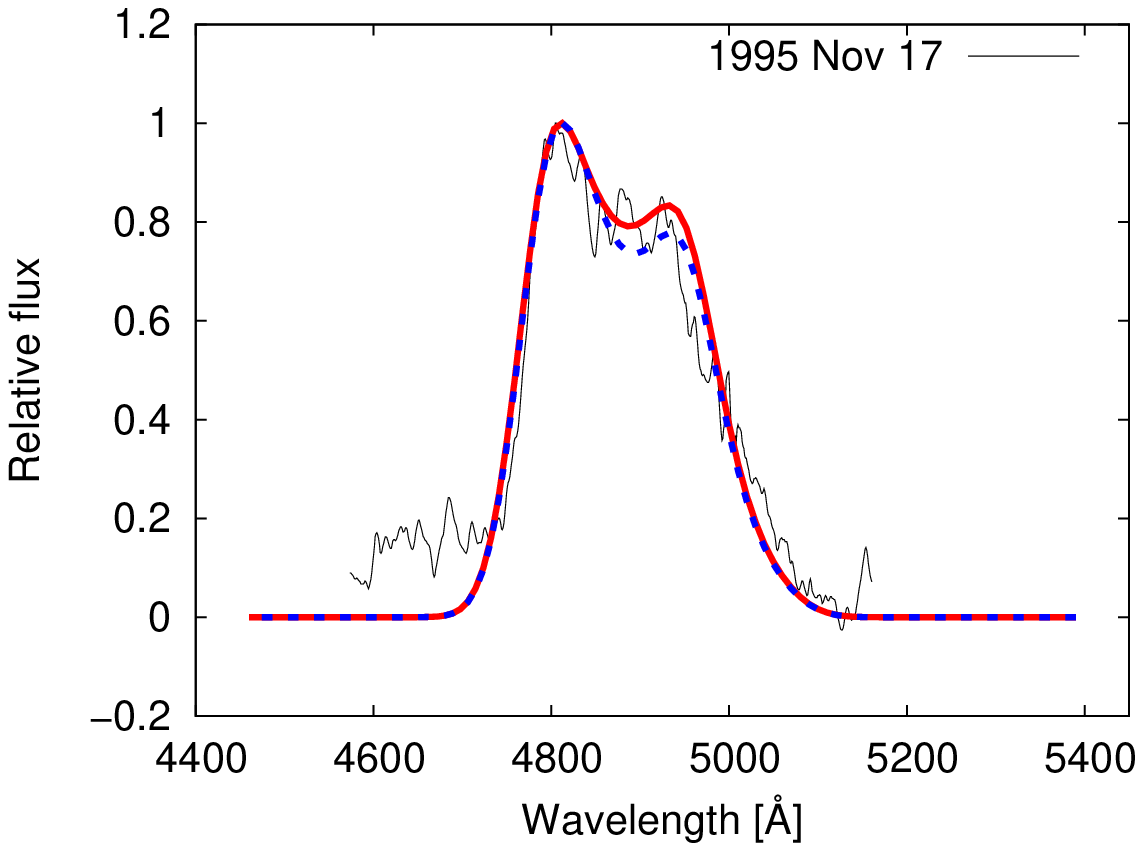}
\includegraphics[width=0.32\textwidth]{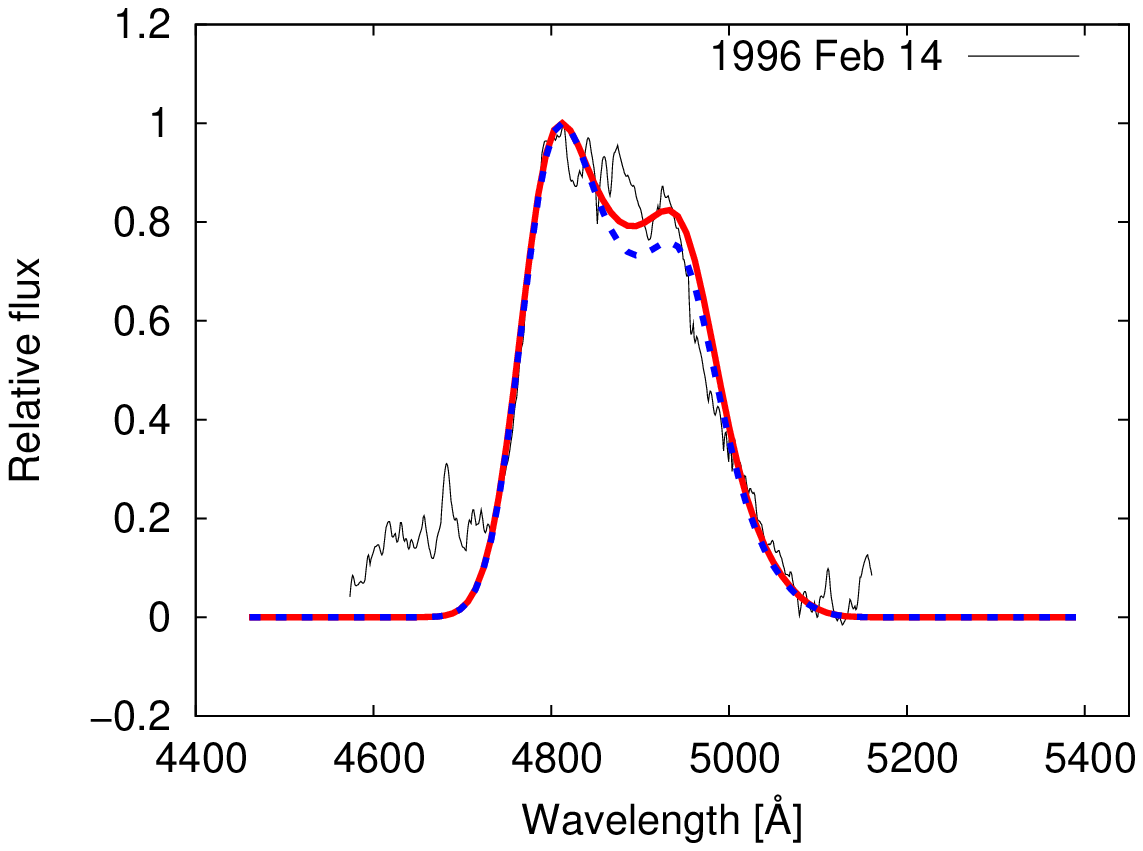}
\includegraphics[width=0.32\textwidth]{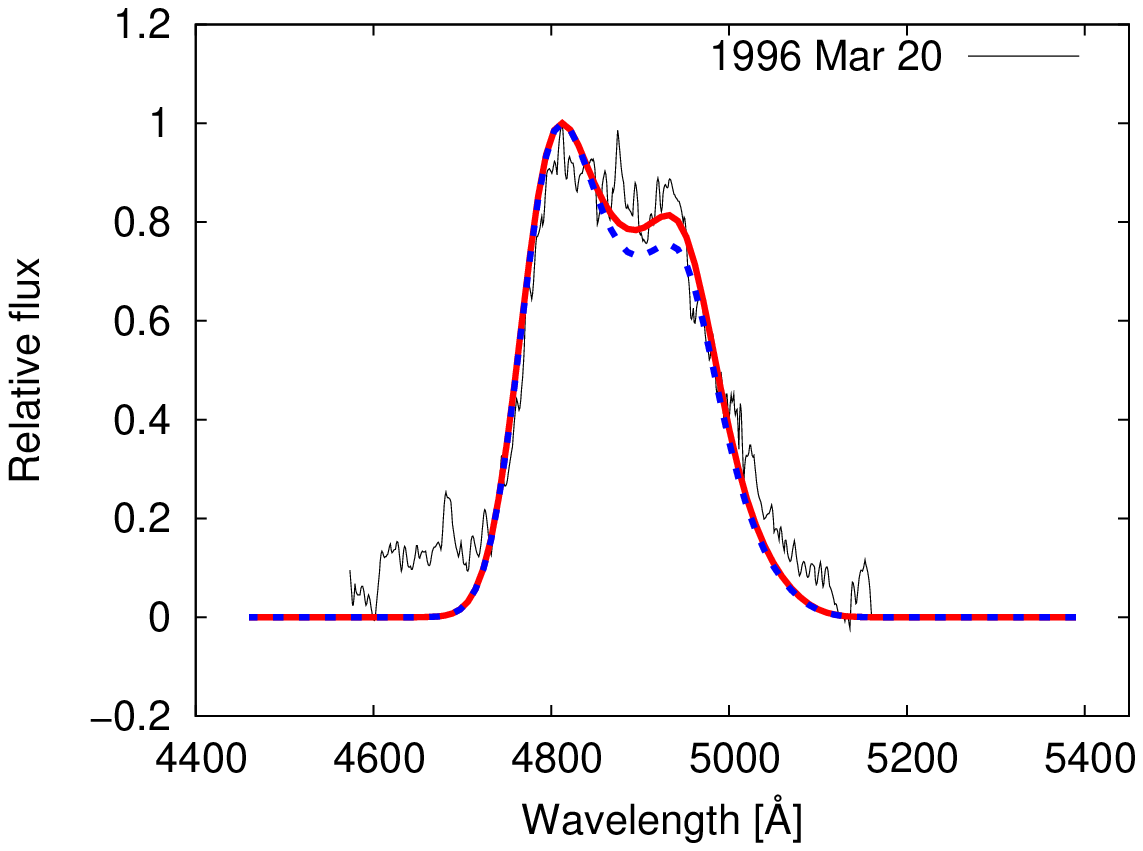} \\
\includegraphics[width=0.32\textwidth]{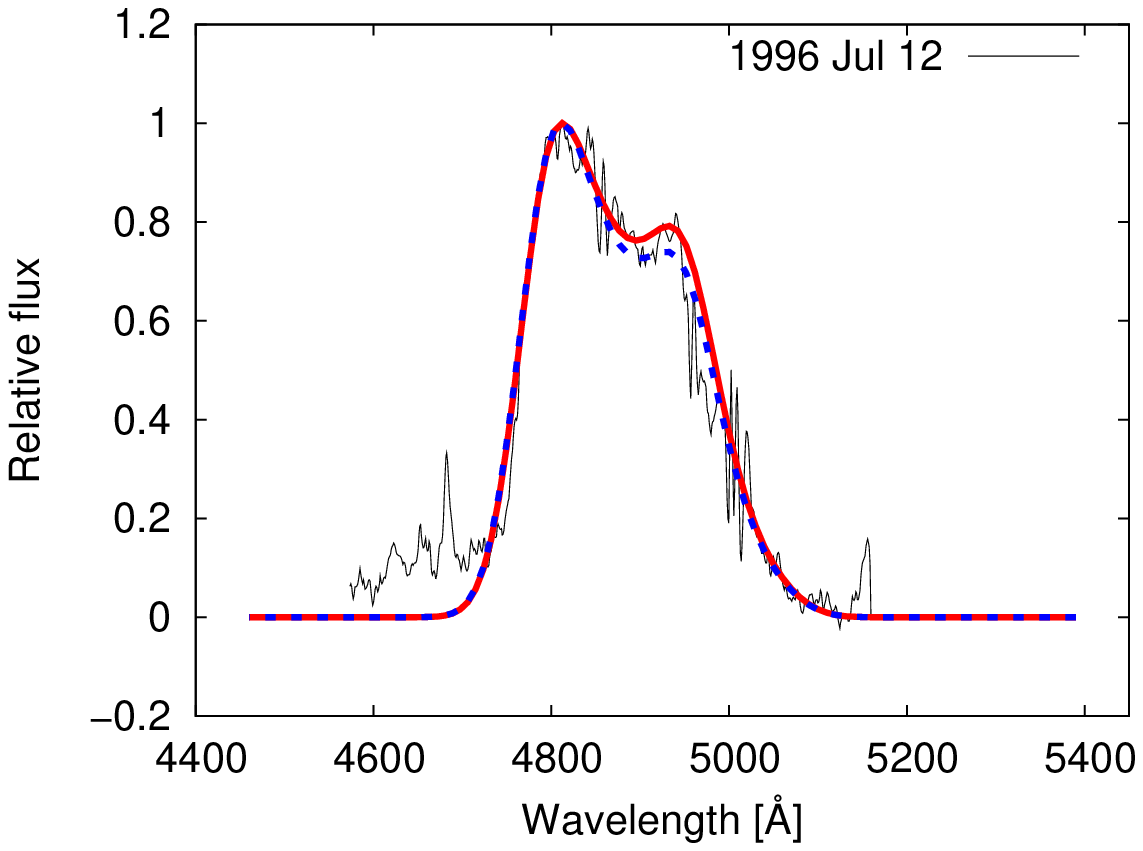}
\includegraphics[width=0.32\textwidth]{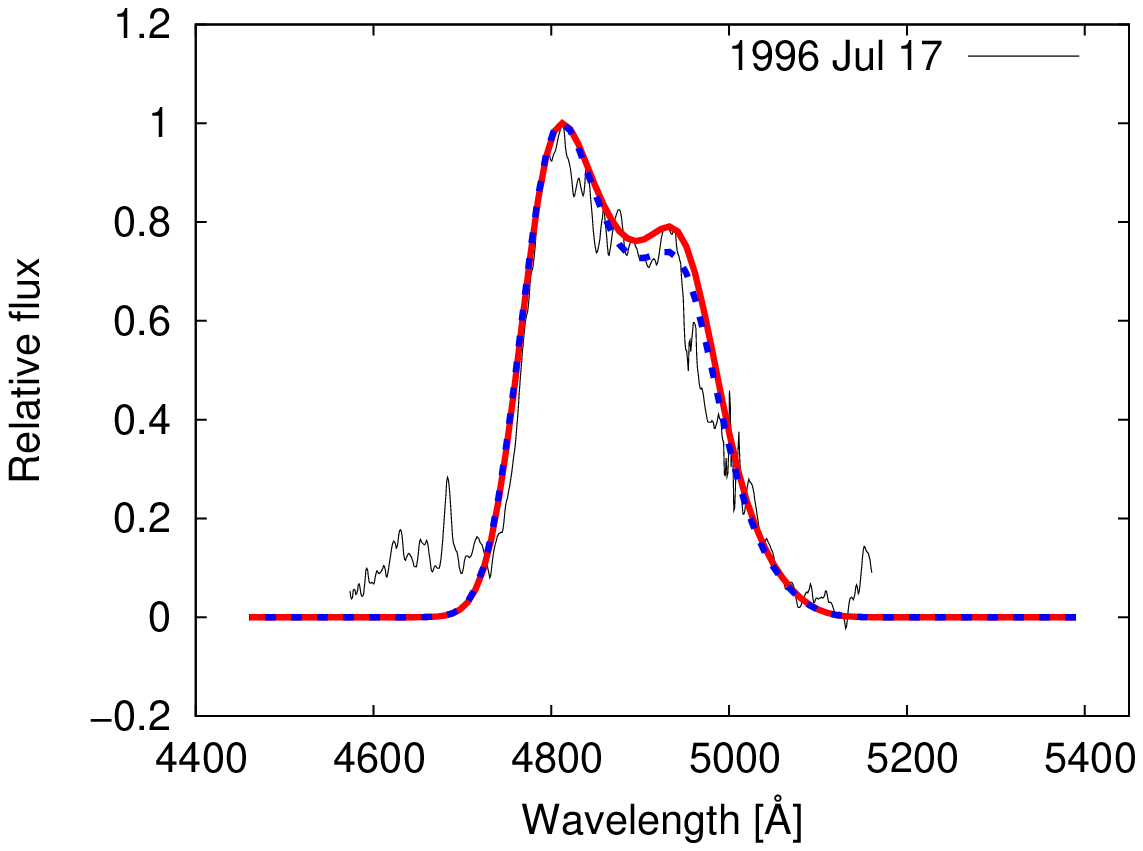}
\includegraphics[width=0.32\textwidth]{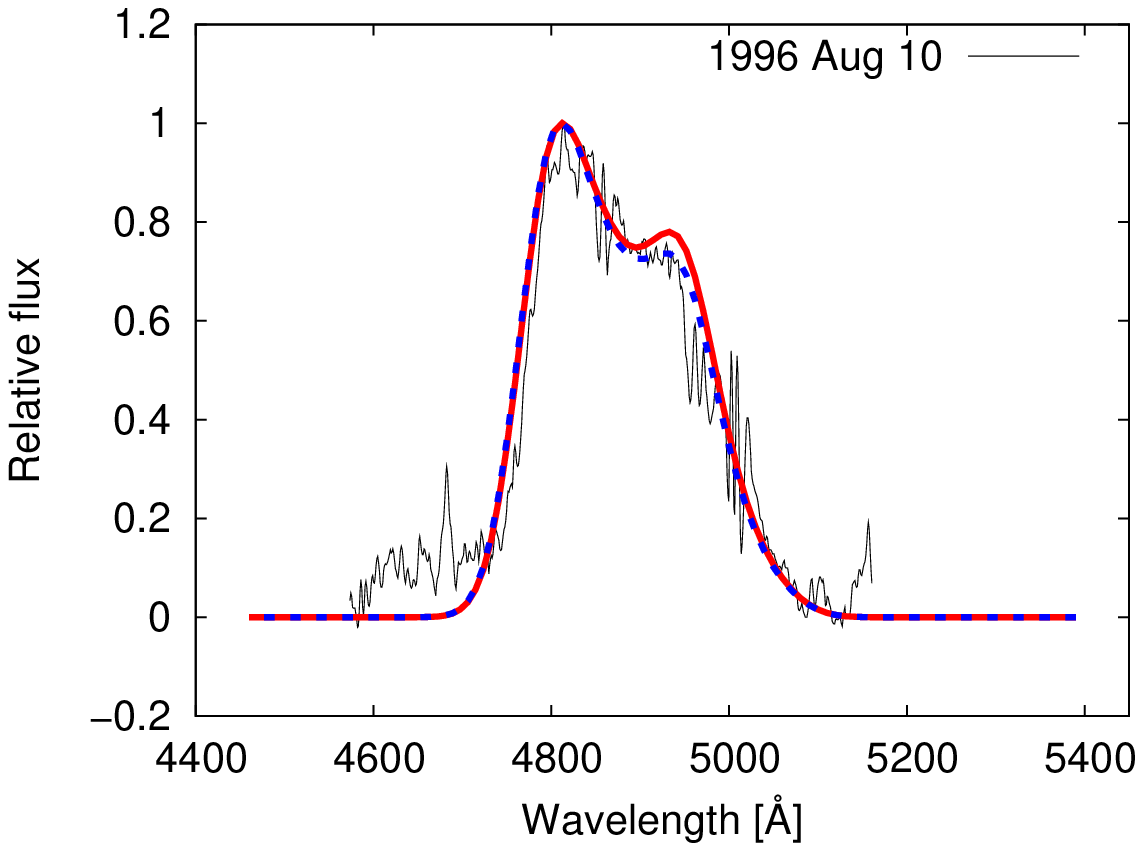} \\
\includegraphics[width=0.32\textwidth]{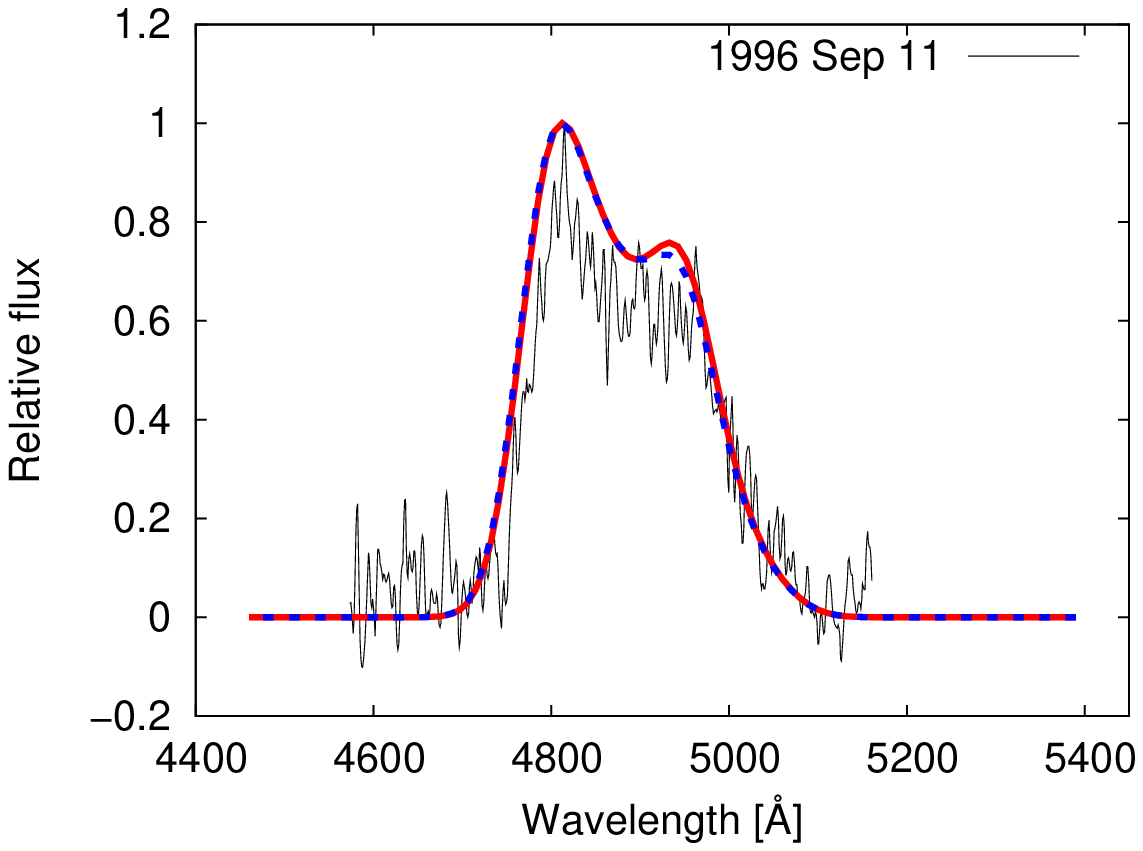}
\includegraphics[width=0.32\textwidth]{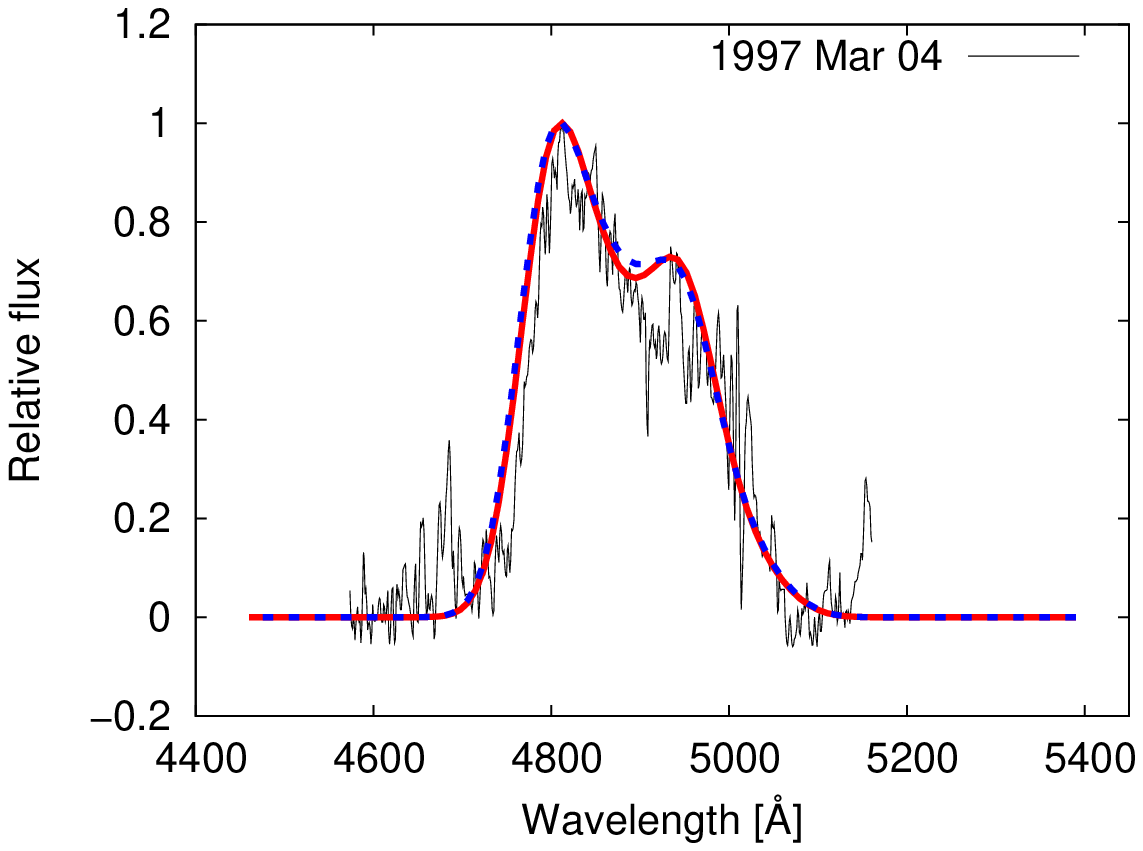}
\includegraphics[width=0.32\textwidth]{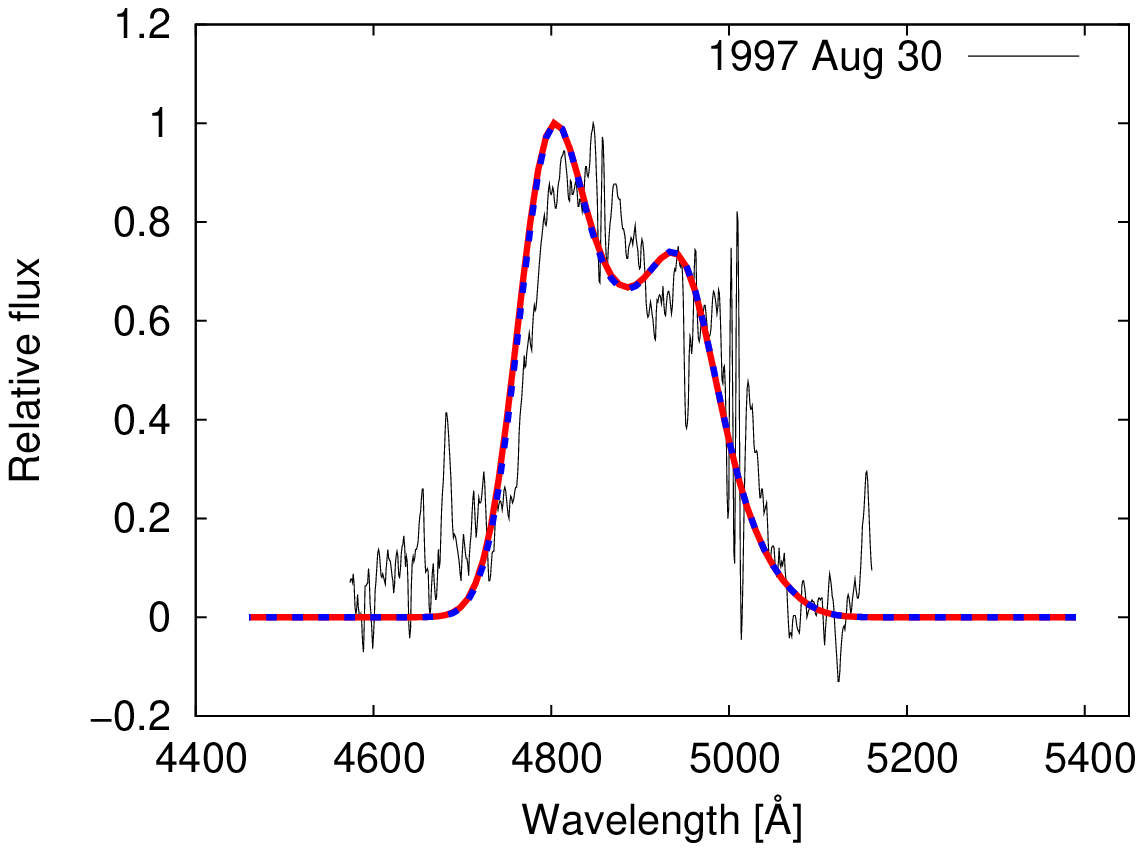} \\
\includegraphics[width=0.32\textwidth]{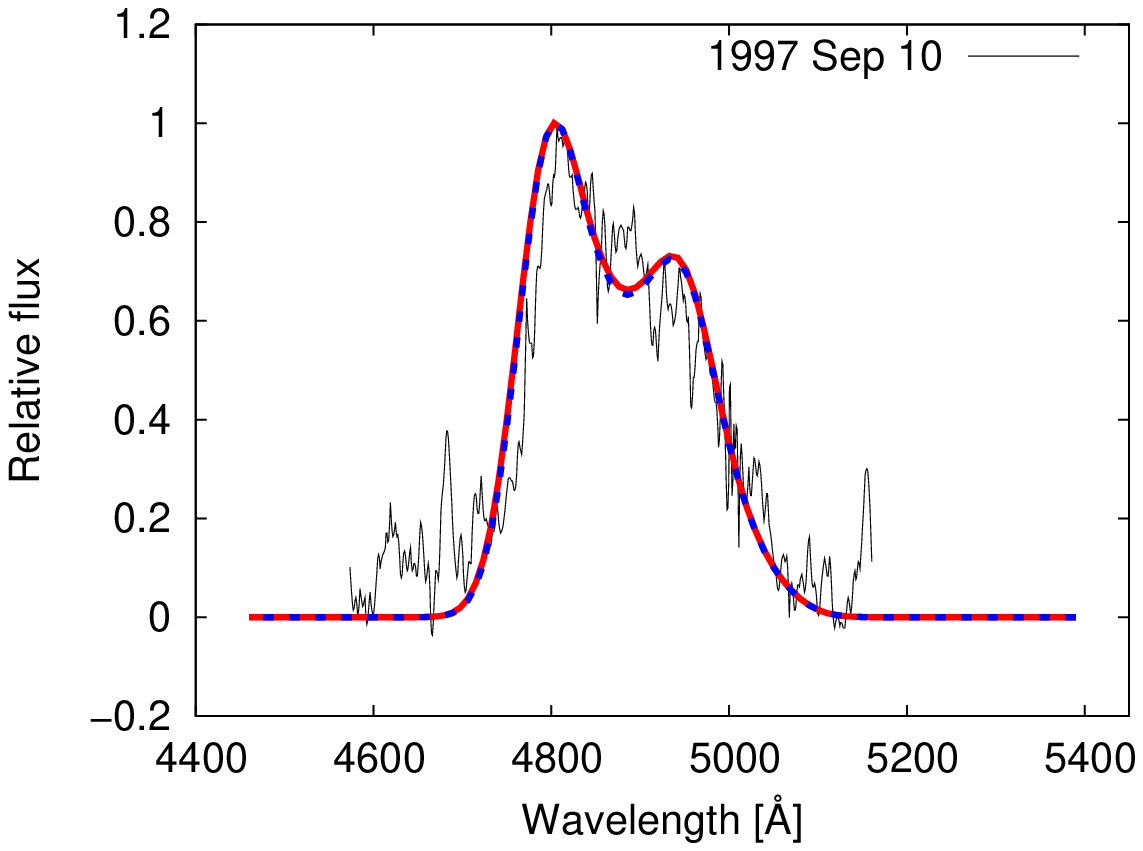}
\includegraphics[width=0.32\textwidth]{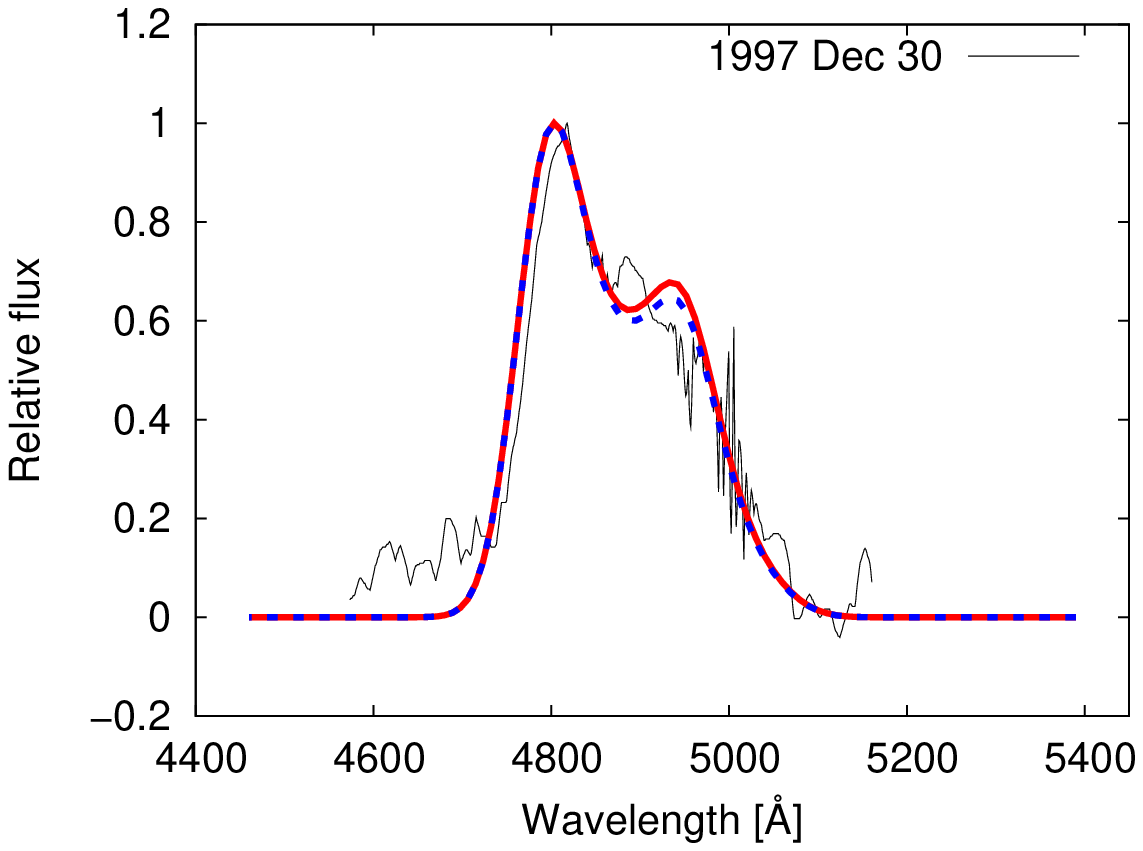}
\includegraphics[width=0.32\textwidth]{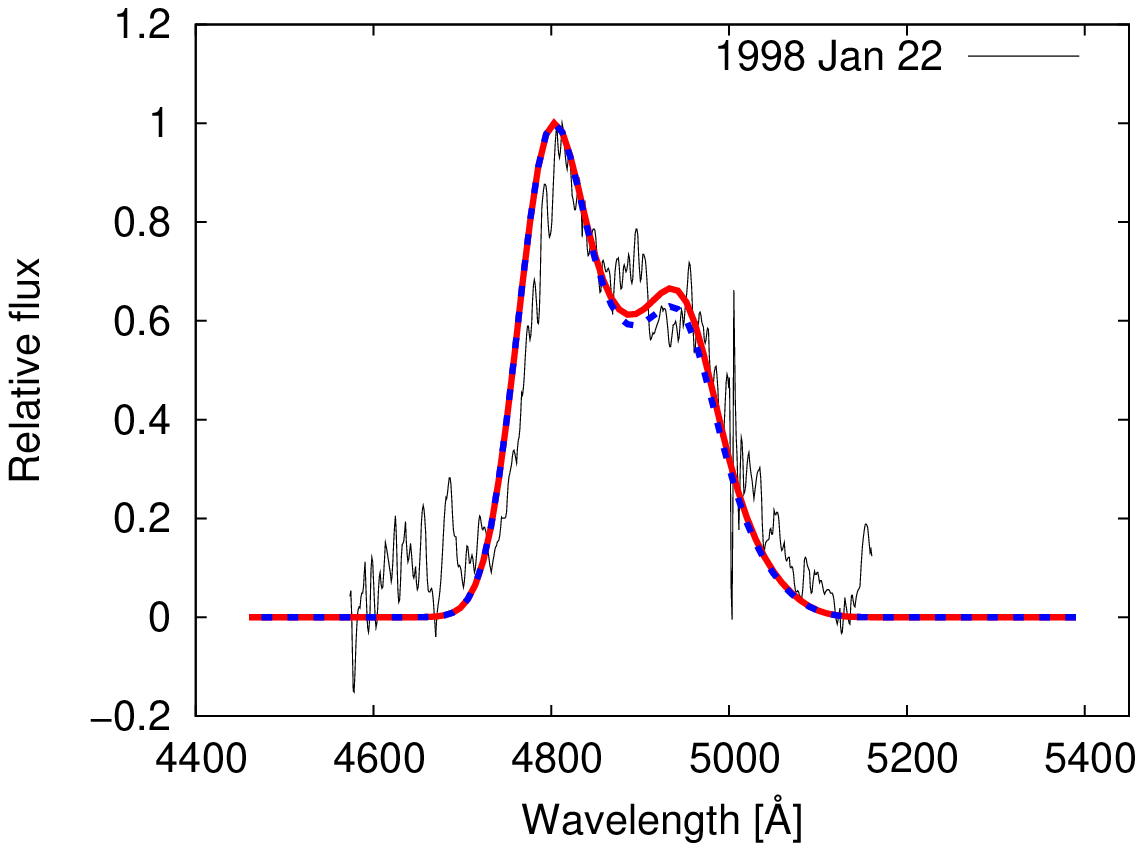}
\caption{Comparisons between the observed H$\beta$ line profiles of quasar 3C 390.3 (black solid line)
and the corresponding simulated profiles due to two successive bright spots. The red solid line represents the simulated
profiles due to the moving bright spots which estimated positions are presented in Fig. \ref{fig5} and widths in Table
\ref{tab1}. The blue dashed line corresponds to the simulated profiles due to the stationary bright spots, positioned
at $x=-100\ R_g$, $y=220\ R_g$ during the first outburst and at $x=-220\ R_g$, $y=125\ R_g$ during the second
outburst.}
\label{fig4}
\end{figure*}


\addtocounter{figure}{-1}
\begin{figure*}
\centering
\includegraphics[width=0.32\textwidth]{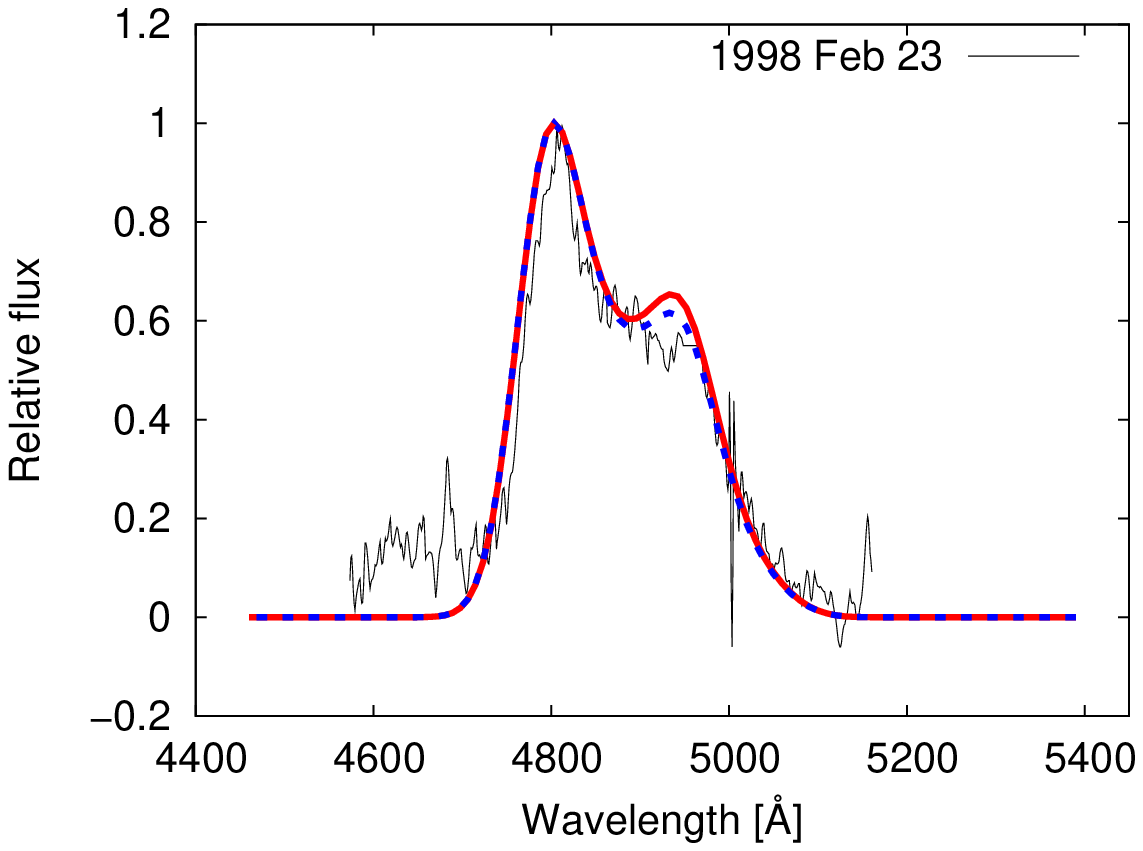}
\includegraphics[width=0.32\textwidth]{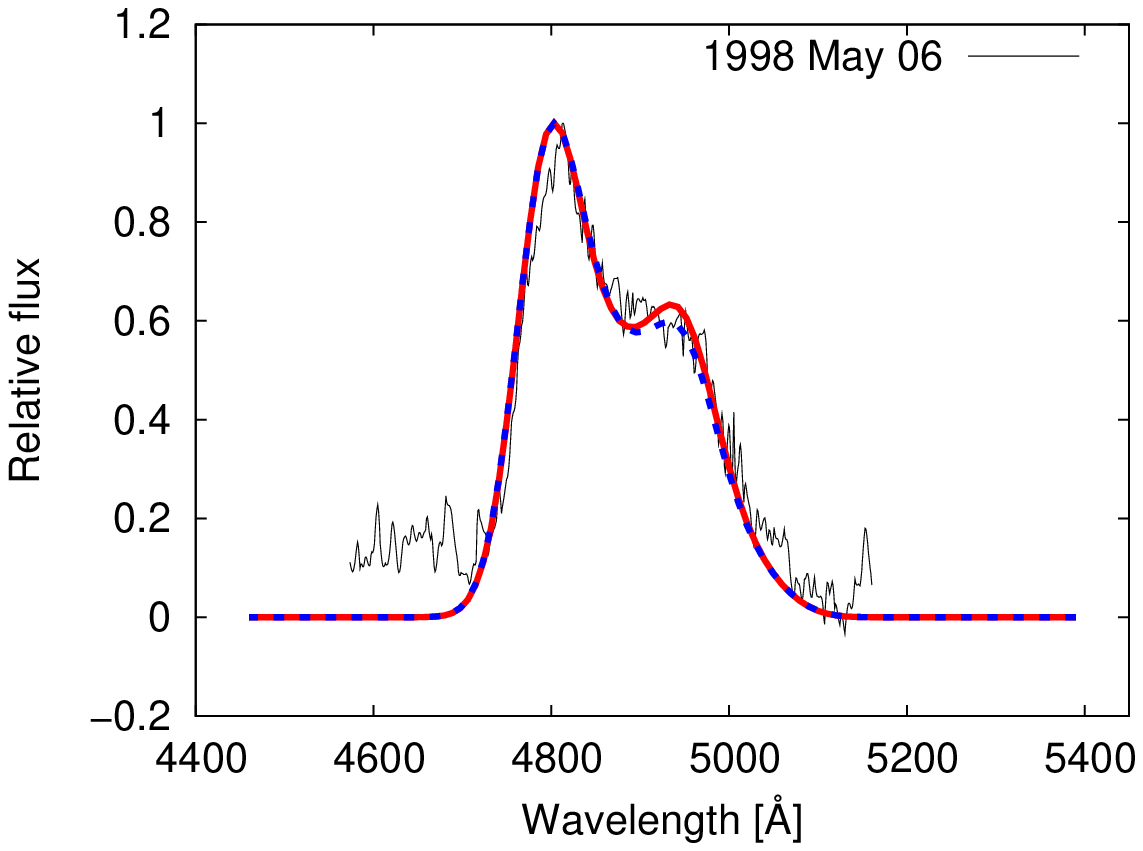}
\includegraphics[width=0.32\textwidth]{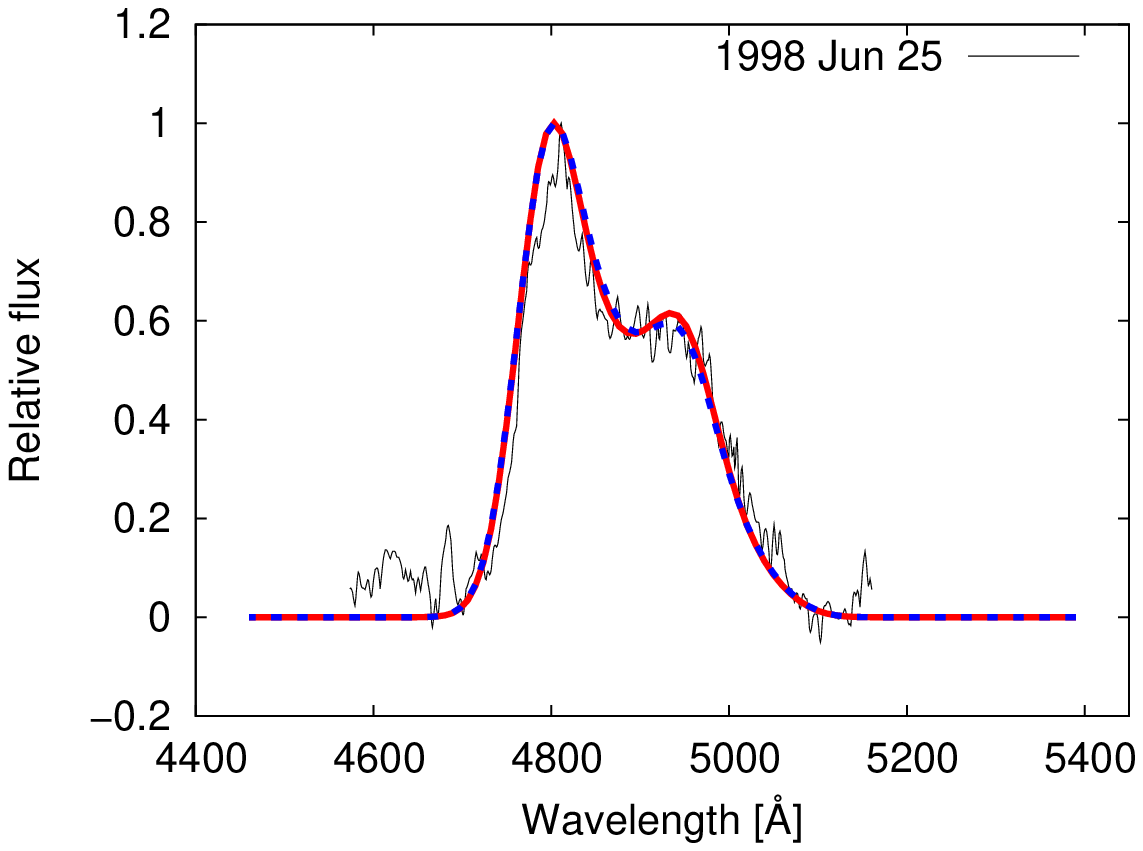} \\
\includegraphics[width=0.32\textwidth]{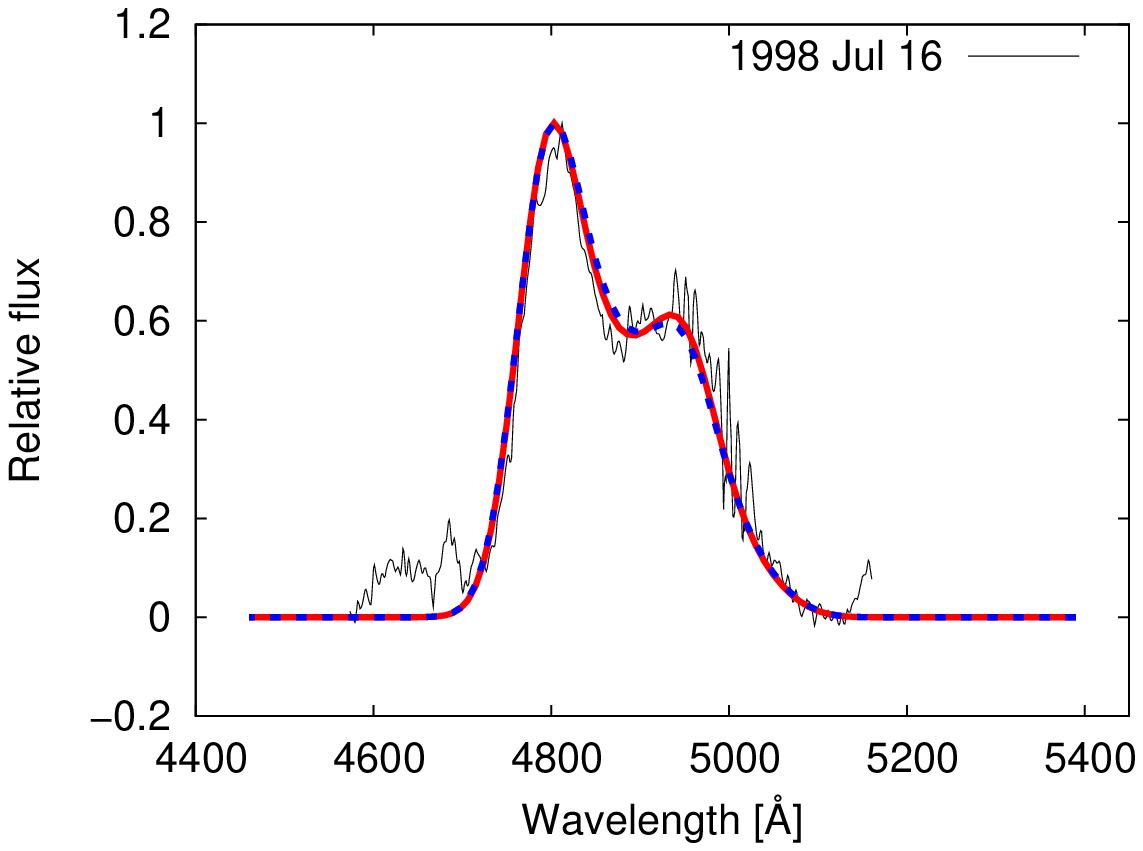}
\includegraphics[width=0.32\textwidth]{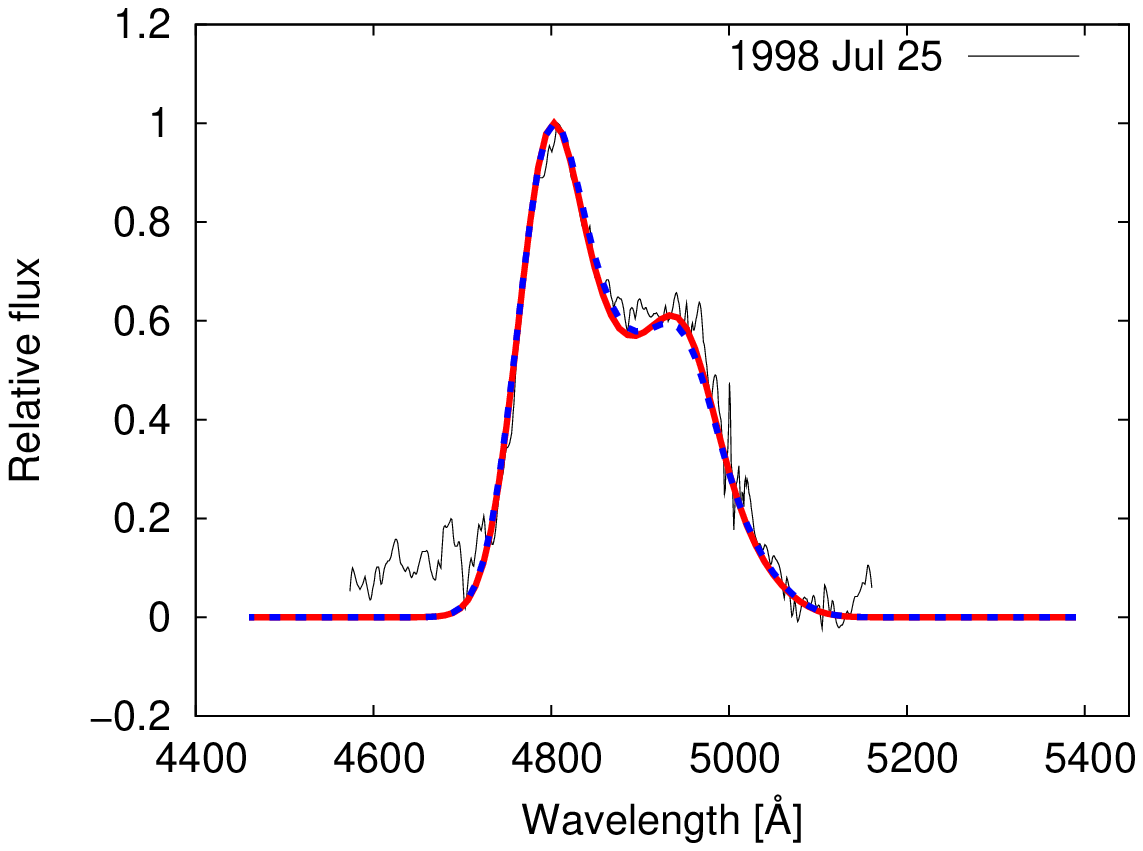}
\includegraphics[width=0.32\textwidth]{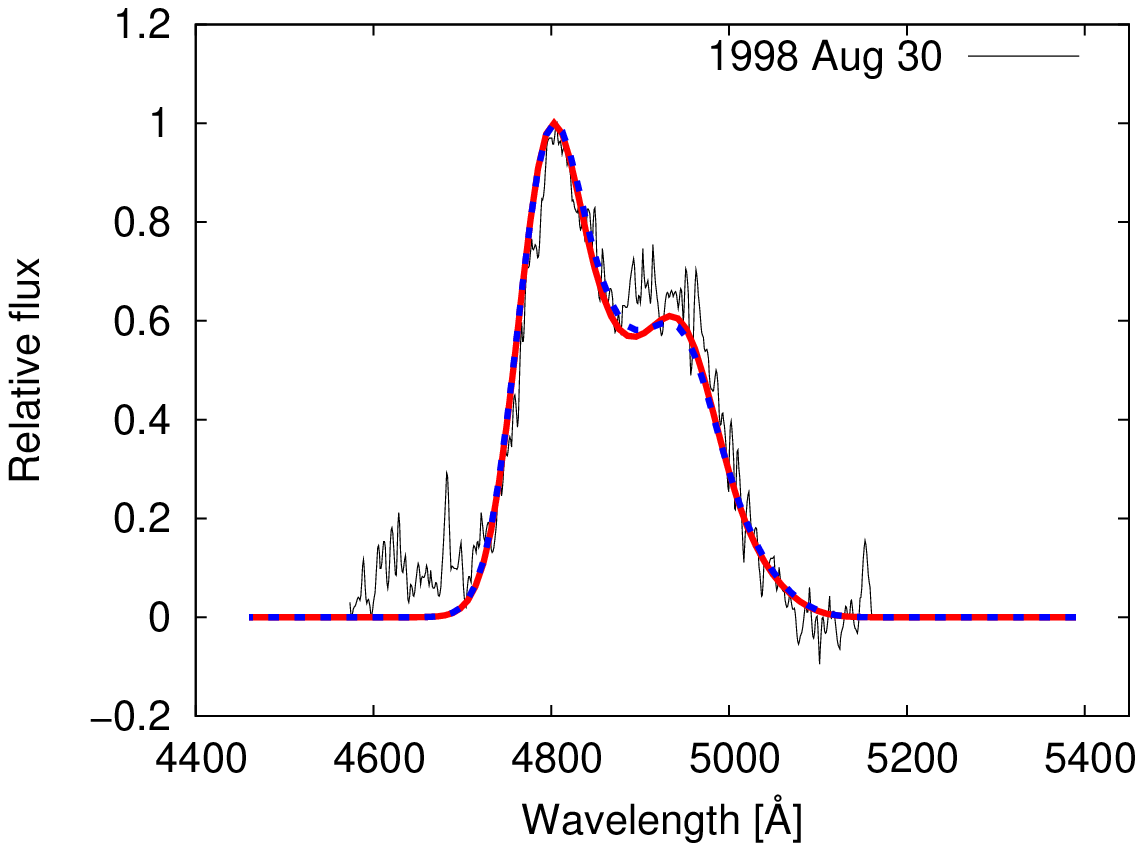} \\
\includegraphics[width=0.32\textwidth]{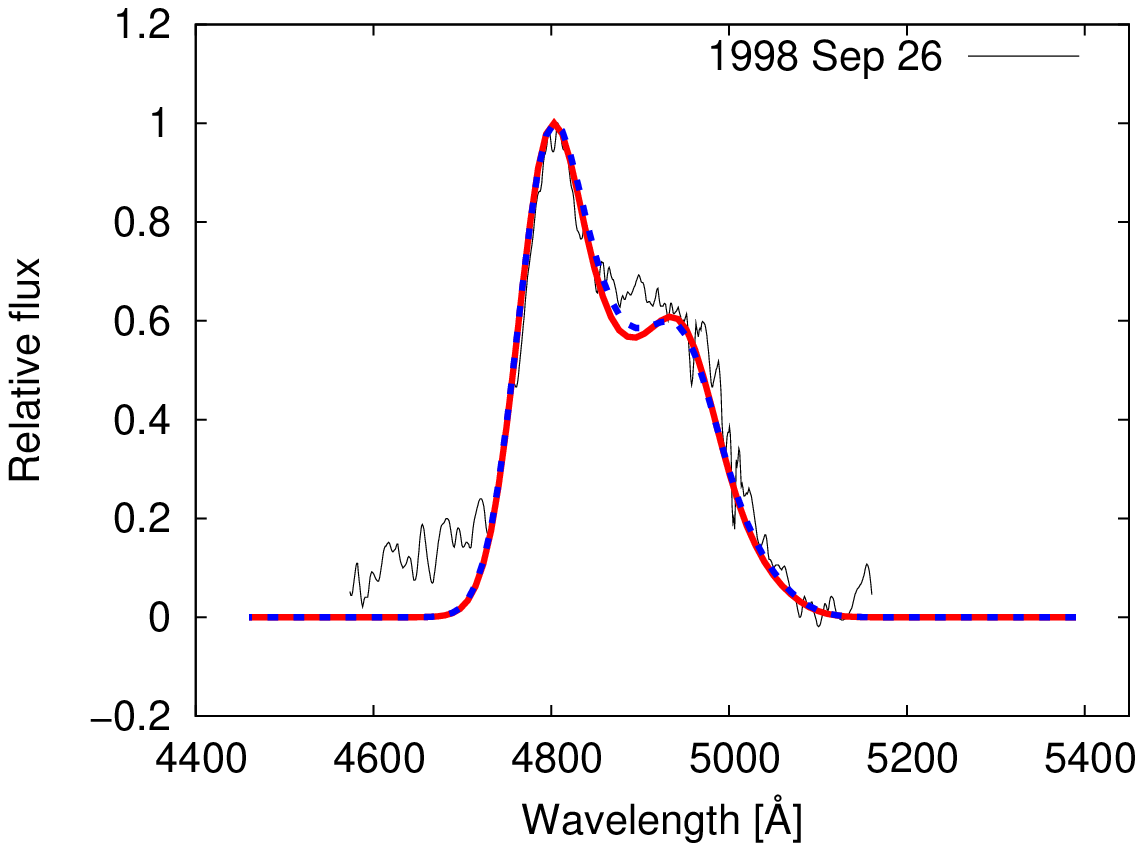}
\includegraphics[width=0.32\textwidth]{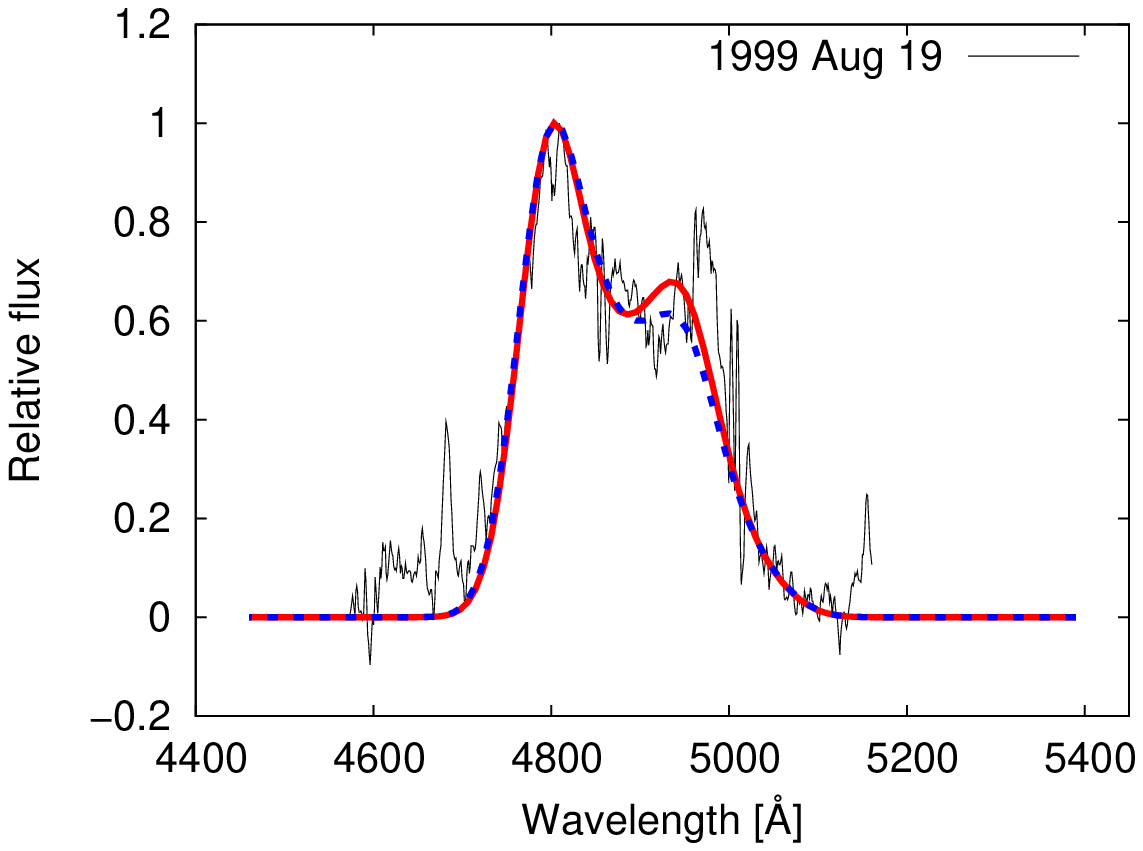}
\includegraphics[width=0.32\textwidth]{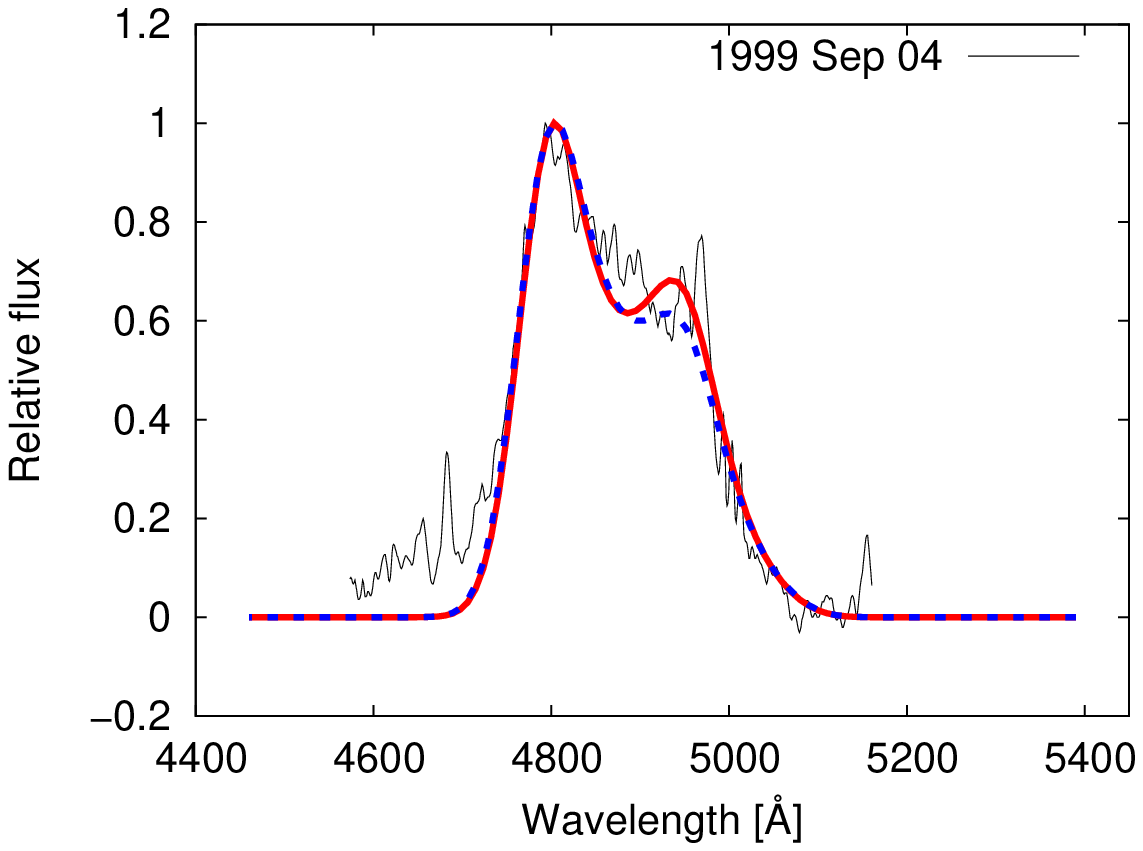} \\
\includegraphics[width=0.32\textwidth]{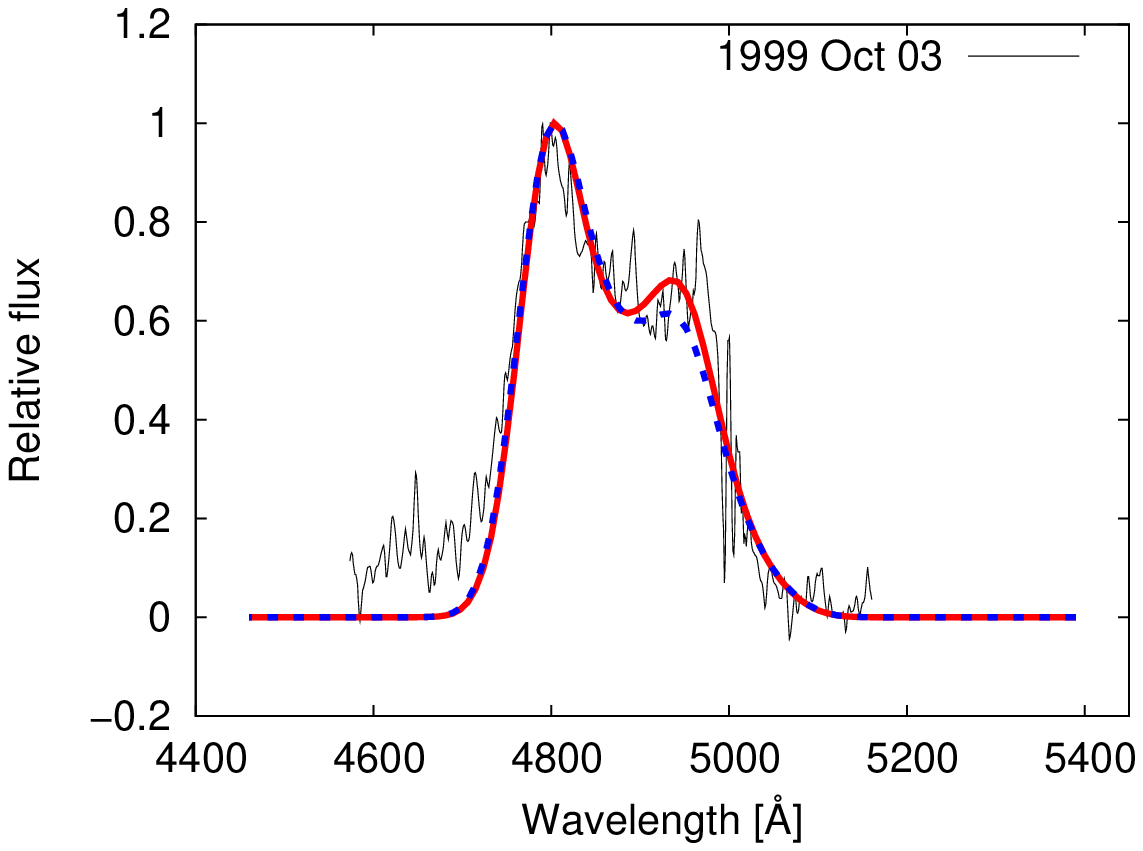}
\caption{(continued)}
\label{fig4b}
\end{figure*}

\begin{figure*}
\centering
\includegraphics[width=0.65\textwidth]{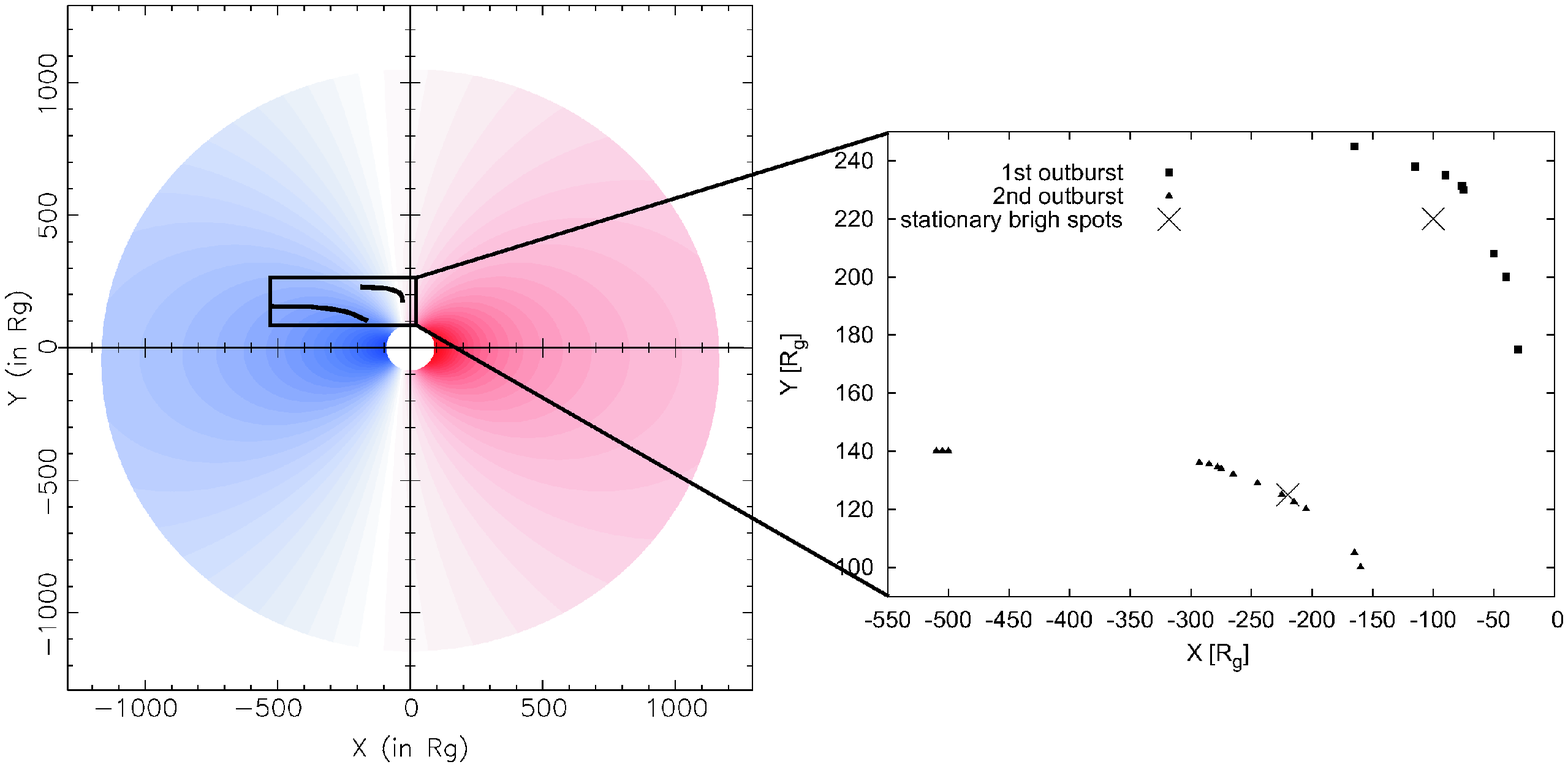}
\caption{Positions of moving perturbing region along the accretion disk corresponding
to two observed amplitude outbursts: October 1994 -- July 1997 outburst
(squares) and July 1997 -- June 1999 outburst (triangles). In both cases, the moving
perturbation originates in the inner regions of the disk and spirals away towards
its outer parts. For average speed of the first perturbation we obtained value of
7298 $\mathrm{km\ s^{-1}}$ and for the second one 6575 $\mathrm{km\ s^{-1}}$. Positions of the
stationary perturbing region (denoted by crosses in the right panel) are: $x=-100\ R_g$, $y=220\ R_g$
during the first outburst and $x=-220\ R_g$, $y=125\ R_g$ during the second outburst.}
\label{fig5}
\end{figure*}

Fig. \ref{fig4}. shows comparisons between all 22 observed spectra (black solid line) and
the obtained best fits assuming the moving (red solid line) and stationary
(blue dashed line) perturbing regions, while the
corresponding positions of the perturbing regions are presented in Fig.
\ref{fig5}. As it can be seen from Fig. \ref{fig5}, the obtained positions of
the moving perturbing regions are distributed in the form of two spiral arms, indicating
that these perturbations originate in the inner regions of the disk and spiral away
towards its outer parts, moving faster in the azimuthal direction as they get
further away from the center of the disk.
Table \ref{tab1}. contains the obtained values of the fitted parameters, as well as the
corresponding RMSD in the case of both, moving and stationary perturbing regions.
We first studied the moving perturbations, but after examining the obtained results we
found that, during a period of a few years, such perturbations move only by small distances
which are comparable to their widths (see Table \ref{tab1}). Therefore, in order to test
whether the obtained displacements are reliable, we repeated the fitting, but this time
assuming the stationary perturbing regions with variable widths.
In this case we obtained the best fits for two perturbations
positioned at $x=-100\ R_g$, $y=220\ R_g$ and at $x=-220\ R_g$, $y=125\ R_g$, respectively
(denoted by crosses in Fig. \ref{fig5}).
As one can see from Fig. \ref{fig4}. and Table \ref{tab1}, regardless the significant
variations of the H$\beta$ line profile during the analyzed period, both models of perturbing region
resulted in similar fits for the most of these spectra, except for the spectra observed during 1999,
where the moving perturbing region achieved better fits. The latter result should be taken with caution
because the observations from 1999 were performed after a large gap of $\approx$ 1 year.
Therefore, the displacements of the perturbing regions cannot be considered as
indisputably confirmed, but on the contrary, their widths almost certainly vary with time.
The last conclusion is valid for both, moving and stationary perturbing regions, since neither
of them can provide satisfactory fits with fixed widths.

The obtained best fit positions of both, moving and stationary perturbations are located
on the approaching side of the accretion disk (see Fig. \ref{fig5}) and these
perturbations can be most likely attributed to successive occurrences
of two different bright spots. This assumption is in good agreement with observations,
since two large amplitude outbursts of H$\beta$ line are observed during
the analyzed period \citep{sh01}, and therefore each bright spot can
be assigned to one of them: the bright spot which positions are
denoted by squares corresponds to October 1994 - July 1997
outburst  while the other one, which positions are denoted by triangles,
corresponds to July 1997 - June 1999 outburst. Using the
time differences between two successive observed spectra we were
able to estimate the speeds of both moving bright spots (see Table
\ref{tab1}). For an average velocity of the first bright spot we
obtained the value of 7298 $\mathrm{km\ s^{-1}}$ and for the second one 6575
$\mathrm{km\ s^{-1}}$. As it can be seen from Table \ref{tab1}, widths of bright
spots are increasing with time, indicating that they decay until they completely disappear.
It should be noted that, inevitably, there is a certain degree of
degeneracy in the parameter space, since similar results could be
obtained with somewhat different combinations of perturbing region
positions and widths.

\begin{deluxetable}{lccccccccc}
\tabletypesize{\small}
\tablecolumns{10}
\tablecaption{Parameters of perturbing region obtained by fitting the observed spectra.
\label{tab1}}
\tablewidth{0pt}
\tablehead{
\colhead{Date} & \colhead{JD (2400000+)} & \colhead{$x [R_g]$} &
\colhead{$y [R_g]$} & \multicolumn{2}{c}{$w [R_g]$} & \colhead{$d [R_g]$} &
\colhead{$v [\mathrm{km\ s^{-1}}]$}& \multicolumn{2}{c}{RMSD} \\
\colhead{(1)} & \colhead{(2)} & \colhead{(3)} & \colhead{(4)} & \colhead{(5)} &
\colhead{(6)} & \colhead{(7)} & \colhead{(8)} & \colhead{(9)} & \colhead{(10)}
}
\startdata
\sidehead{The First Outburst}
1995 Nov 17	& 50039.156& -30   & 175   & 100  & 100 &     &	       &0.09730 &0.09460 \\
1996 Feb 14	& 50127.602& -40   & 200   & 105  & 130 &26.93& 5203.24&0.09680 &0.09900 \\
1996 Mar 20	& 50162.580& -50   & 208   & 106.5& 140 &12.81& 6257.63&0.08476 &0.09138 \\
1996 Jul 12	& 50276.567& -75   & 230   & 110  & 180 &33.3 & 4993.36&0.08518 &0.08178 \\
1996 Jul 17	& 50281.434& -76.3 & 231.3 & 110  & 180 & 1.84& 6456.24&0.08820 &0.08249 \\
1996 Aug 10	& 50305.489& -90   & 235   & 112  & 190 &14.19&10082.9 &0.09435 &0.09201 \\
1996 Sep 11	& 50338.309& -115  & 238   & 120.5& 200 &25.18&13112.6 &0.12337 &0.12426 \\
1997 Mar 04	& 50511.622& -165  & 245   & 130  & 260 &50.49& 4978.94&0.11448 &0.12082 \\
\sidehead{The Second Outburst}
1997 Aug 30	& 50691.463& -160  & 100   & 100  & 100 &	    &	 &0.13717 &0.13593 \\
1997 Sep 10	& 50701.576& -165  & 105   & 105  & 110 & 7.07 &11950.55 &0.12034 &0.11905 \\
1997 Dec 30	& 50813.195& -205  & 120   & 145  & 175 &42.72 &6541.48  &0.09499 &0.09398 \\
1998 Jan 22	& 50835.631& -215  & 122.5 & 155  & 195 &10.31 &7852.39  &0.10690 &0.10949 \\
1998 Feb 23	& 50867.560& -225  & 125   & 165  & 215 &10.31 &5517.75  &0.09825 &0.09618 \\
1998 May 06	& 50940.354& -245  & 129   & 185  & 265 &20.40 &4788.88  &0.08867 &0.08965 \\
1998 Jun 25	& 50990.302& -265  & 132   & 205  & 285 &20.22 &6920.32  &0.07229 &0.07387 \\
1998 Jul 16	& 51010.719& -275  & 134   & 210  & 295 &10.20 &8537.04  &0.06809 &0.07141 \\
1998 Jul 25	& 51019.723& -278  & 134.5 & 212  & 300 & 3.04 &5773.22  &0.06548 &0.06615 \\
1998 Aug 30	& 51055.551& -285  & 135.5 & 215  & 320 & 7.07 &3373.22  &0.07590 &0.07638 \\
1998 Sep 26	& 51082.429& -293  & 136   & 218  & 340 & 8.01 &5097.09  &0.07948 &0.07788 \\
1999 Aug 19	& 51410.309& -500  & 140   & 250  & 400 &207.04&10792.43 &0.10645 &0.11163 \\
1999 Sep 04	& 51426.208& -505  & 140   & 252  & 400 &  5   &5375.06  &0.09147 &0.09232 \\
1999 Oct 03	& 51455.172& -510  & 140   & 254  & 400 &  5   &2950.49  &0.09977 &0.10270 \\
\enddata
\tablecomments{Col. (1): date of observation; Col. (2): epoch of observation (in JD);
Cols. (3-4): $x$ and $y$ coordinates of the moving bright spots;
Col. (5): widths ($w=w_x=w_y$) of the moving bright spots;
Col. (6): widths ($w=w_x=w_y$) of the stationary bright spots, positioned at
$x=-100\ R_g$, $y=220\ R_g$ during the first outburst and at $x=-220\ R_g$, $y=125\ R_g$ during the second outburst;
Col. (7): linear distance crossed by the moving bright spots between two successive observations;
Col. (8): average speed of the moving bright spots between two successive observations;
Col. (9): root mean square deviation between the observed and fitted H$\beta$ line profiles in the case of
the moving bright spots;
Col. (10): the same as Col. (9), but in the case of the stationary bright spots.}
\end{deluxetable}

\section{Discussion}

Several physical mechanisms could be responsible for perturbations
in accretion disk emissivity, i.e. for bright spot formations. The
most plausible candidates for such mechanisms are: disk
self-gravity, baroclinic vorticity, disk-star collisions
\citep[][and references therein]{fe08}, tidal disruptions of stars
by central black hole \citep[][and references therein]{sq09}
and fragmented spiral arms \citep[][and references
therein]{le10}.

The disk self-gravity, driven by Jeans instability, could cause
production of clumps in the disk which have typical sizes in the
range from 10 to 1000 gravitational radii for a $10^8\ M_\odot$
central black hole. Such clumps do not shear with differential
rotation and they have high brightness that varies very little over
time. Since the obtained results indicate that bright spots decay by
time and spiral along the disk, it is not likely that these
bright spots could be identified as clumps created by self-gravity,
although their sizes are comparable.

Baroclinic vorticity appears in the accretion disk due to its differential
rotation in combination with the radial temperature gradient, causing the
material in the disk to spiral around the center of vortex. Such a
vortex would have higher density, and hence higher brightness, causing the
formation of a bright spot. The typical sizes of such spots, as well as their
shearing with differential rotation of the disk, are still unknown since
different numerical simulations gave contradictory results \citep[for more
details see e.g.][and references therein]{fe08}. Therefore, in the case of 3C
390.3, this mechanism still cannot be neither ruled out, nor accepted.

Disk-star collisions are assumed to be very frequent events which happen
on daily timescales and which could increase disk surface temperature in
the region of collision, and thus, create a bright spot. Such bright spots
shear with differential rotation of the disk and decay as the material cools
down. However, the typical size of a such bright spot immediately after
collision is close to the size of the star, which is very small when expressed
in gravitational radii. Therefore, neither this mechanism could be
accepted as a potential cause of two bright spots, detected in the case of
quasar 3C 390.3.

The tidal disruption of stars by central black hole
\citep[][and references therein]{sq09} happens when a star passes
the tidal radius of the black hole, i.e. when the black hole's tidal
gravity exceeds the star's self-gravity. Gas of a disrupted star
falls back to the black hole at a super-Eddington rate, releasing a
flare of energy which then blows away a significant fraction of the
falling gas as an outflow. Such super-Eddington flares and outflows
could induce instabilities in accretion disk in form of bright
spots. However, this mechanism is an unlikely candidate for a
potential cause of the bright spots in the case of 3C 390.3 due to
the following weaknesses: (i) super-Eddington outflows are short
living ($\sim 10$ days); (ii) frequency of star disruptions in a
typical elliptical galaxy is very low, between $10^{-5}$ and
$10^{-4}$ per year, and in the case of the black hole of 3C 390.3
which mass is $\sim 5 \times 10^8 M_\odot$ \citep{le06}, it is near
the low end of this range \citep[see e.g.][]{wa04,mt99}; (iii) any
stellar debris could be hardly released in the case of 3C 390.3
since the main sequence stars are disrupted within the innermost
stable orbit around a non-rotating black hole as massive as that in
3C 390.3; (iv) it is unclear how the bright spots produced by this
mechanism should evolve with time.

Spiral arms in AGN disks could be formed spontaneously due
to self-gravity instabilities \citep[see e.g.][and references
therein]{fe08} or could be triggered by close passage of some
massive object such as another supermassive black hole or a star
cluster \citep[see e.g.][and references therein]{le10}. Spiral arms
increase flux variability of AGN on timescales of a year to several
years, but as noted before, they are unable to reproduce all aspects
of the observed variability. However, they are also subjected to
fragmentation, causing small variations in the flux on timescales of
several months. The fragments in spiral arms can be due to
sub-structures in a non-uniform accretion disk, such as isolated
clumps which could pass through the arm and dominate in its
emissivity, causing the discrete "lumps" of excess emission
\citep{le10}. The observed variability on timescales from few months
to several years in the difference spectra of some AGN is probably
caused by such lumps. It is quite possible that some of these lumps
are long-living and that they do not vary significantly in strength,
shape, or position over a period of several years \citep{le10}. As
the obtained results show, it is the same case with the large bright
spots which are responsible for amplitude outbursts of the 3C 390.3
H$\beta$ line, because they have constant emissivity and they are
either stationary or spiralling over small distances during the
period of several years. The only feature which significantly varies
with time is their width. Therefore, these bright spots could be
most likely explained by the emissivity lumps, caused by fragments
in spiral arms of the accretion disk.

\section{Conclusions}

We developed a model of the disk perturbing region in the form of a
single bright spot (or flare) by a modification of the power law
disk emissivity and used this model to simulate the disk line
profiles. This model has been used to fit the observed H$\beta$ line
of 3C 390.3 observed from 1995 to 1999. From this investigation we
can point out the following results:

\begin{enumerate}
\item The model which includes perturbation (bright spot) in the
    accretion disk can successfully explain difference in double
    peaked line profiles, as e.g. higher red peak even if we
    have the standard circular disk. The position of a bright
    spot has a stronger influence on one particular part of
    spectral line profiles (such as e.g. its core if the spot is
    in the central part of the disk, or "red" and "blue" wings
    if the spot is located on receding and approaching part,
    respectively).
\item Using the model for perturbing region we were able to
    successfully model and reproduce the observed variations of
    the H$\beta$ line profile in the case of 3C 390.3, including
    the two large amplitude outbursts observed during the
    analyzed period. Therefore, the observed variations of the 3C 390.3
    H$\beta$ line could be caused by perturbations
    in the disk emissivity.
\item We found that two  outbursts referred by \citet{sh01}
    could be explained by successive occurrences of two
    different bright spots on approaching side of the disk which are either
    moving, originating in the inner regions of the disk and spiralling outwards, or stationary.
    Both bright spots decay by time until they completely disappear.
\item Our results support hypothesis that the perturbations in
    accretion disk emissivity are probably caused by
    fragments in the spiral arms of the disk.
\end{enumerate}

The results presented above show that a circular disk with
perturbations (bright spots) can be applied to explain different
double peaked line profiles, and can be also used to trace
perturbations (as well as their characteristics) from the broad
double peaked line shapes.

\acknowledgments

This work is supported by the Ministry of Science of Serbia through project (146002) “Astrophysical Spectroscopy of
Extragalactic Objects” and with studentship for M. Stalevski. It is also supported by INTAS (grant N96-0328), and
RFBR (grants N97-02-17625, 09-02-01136). The authors would
like to thank the anonymous referee for very valuable and helpful
comments and suggestions.

\end{document}